\documentclass{aa}
\usepackage{graphicx}
\usepackage{epsfig}
\usepackage{longtable,lscape}
\usepackage{natbib}
\usepackage{rotating}
\bibpunct{(}{)}{;}{a}{}{,} 
\begin{document}

\title{An optical/NIR survey of globular clusters in early-type galaxies}
\subtitle{II. Ages of GC systems and the relation to galaxy morphology}

\author{A. L. Chies-Santos\inst{1}, S. S. Larsen\inst{1}, H. Kuntschner\inst{3}, P. Anders\inst{1}, E. M Wehner\inst{1, }\inst{2},  J. Strader\inst{4}, J. P. Brodie\inst{5} and J. F. C. Santos Jr.\inst{6}}

\offprints{A.L.Chies@uu.nl} 

\institute{Astronomical Institute, University of Utrecht, Princetonplein 5, NL-3584, Utrecht, The Netherlands
\and Department of Physics and Astronomy, McMaster University, Hamilton L8S 4M1, Canada 
\and Space Telescope European Coordinating Facility, European Southern Observatory, Garching, Germany
\and Harvard-Smithsonian Center for Astrophysics, Cambridge, MA 02138, USA 
\and UCO/Lick Observatory, University of California, Santa Cruz, CA 95064, USA
\and Departamento de F\'isica, ICEx, Universidade Federal de Minas Gerais, Av. Ant\^onio Carlos 6627, Belo Horizonte 31270-901, MG, Brazil}

\date{Received 3 September 2010 / Accepted 21 October 2010}

\abstract
{Some photometric studies of extragalactic globular cluster (GC) systems using the optical and near-infrared colour combination have suggested the presence of a large fraction of intermediate-age (2-8\,Gyrs) GCs.}
{We investigate the age distributions of GC systems in 14 E/S0 galaxies.}
{We carry out a differential comparison of the $(g-z)$ \textit{vs.} $(g-K)$ two-colour diagrams for GC systems in the different galaxies in order to see whether there are indications of age differences.
We also compare the different GC systems with a few simple stellar population models.}
{No significant difference is detected in the mean ages of GCs among elliptical galaxies. S0 galaxies on the other hand, show evidence for younger GCs.
Surprisingly, this appears to be driven by the more metal-poor clusters.
The age distribution of GCs in NGC\,4365 seems to be similar to that of other large ellipticals (e.g. NGC\,4486, NGC\,4649).
Padova SSPs with recently released isochrones for old ages (14 Gyrs) show less of an offset with respect to the photometry than previously published models.} 
{We suggest that E type galaxies assembled most of their GCs in a shorter and earlier period than S0 type galaxies. The latter galaxy type, seems to have a more extended period of GC formation/assembly.}


\titlerunning{Age distributions of GCs in early-type galaxies}

\authorrunning{Chies-Santos et al.}

\maketitle

\section{Introduction}
Whenever a major star formation burst occurs, globular clusters (GCs) are expected to form along in significant amounts (e.g. \citealt{az92}, \citealt{ee97}, \citealt{fbg97}, \citealt{bs06}).
In the $\Lambda$CDM universe examples of such events that shape the evolution of galaxies are infall of gas into single dark matter halos and (gas rich) mergers.
The latter, are the main mechanism behind the hierarchical merging scenario (e.g. \citealt{white78}, \citealt{springel05}).

GCs still remain the best approximations to simple stellar populations (SSP), despite the recent findings of some clusters hosting multiple stellar populations (e.g. \citealt{piotto07}, in the Milky Way).  
It is difficult to resolve the individual stars of a galaxy beyond the Local Group. On the other hand, GCs can be easily detected at distances as far as the Virgo and Fornax clusters. Hence, they have become a powerful tool in the study of galaxies in the local Universe. 
Observations of GCs translated into for example, ages and metallicities are now widely used to put constraints on the formation and evolution galaxies (\citealt{bs06}).

The importance of dating GCs stems from the possibility of retracing back the major epochs of star formation in a galaxy.
For extragalactic, unresolved GCs, only integrated properties can be measured. Optical colours suffer from the well known age/metallicity degeneracy (\citealt{worthey94}).
This degeneracy can in principle be lifted by combining optical and near-infrared(NIR) colours (e.g. \citealt{puzia02}, \citealt{hempel03}; \citealt{lbs05}).
Generally studies using the optical/NIR colour combination have derived ages by mainly comparing GC colours with SSP models (\citealt{hempel07}, \citealt{hempel07AA}, \citealt{kotulla08}, \citealt{bs06} and references therein).
The typical conclusion found from these studies is that these galaxies host mainly old clusters (with ages $\ga$10\,Gyrs), although non-negligible fractions of younger clusters ($\sim 2-8$ Gyrs) have been claimed in some cases: NGC\,1316, NGC\,4365 and NGC\,5846. 
In NGC\,1316 the presence of intermediate-age ($\sim$2-3\,Gyrs) clusters (\citealt{goud01}) is consistent with its very young luminosity weighted age (\citealt{kunt00}). This implies that NGC\,1316 has experienced a major merger event 3\,Gyrs ago (\citealt{goud01spec}).
However, in NGC\,4365 and NGC\,5846 there is no evidence from integrated light (\citealt{yamada06} and \citealt{thomas05}, respectively) that these galaxies have undergone recent star formation activity.

A general characteristic of GC systems of massive galaxies is having their (optical) colours distributed in a bi(multi)-modal way.
Spectroscopic studies have suggested that colour bimodality is fundamentally due to metallicity differences (e.g. \citealt{strader07}) as most GCs are found to be old (\citealt{strader05}, \citealt{cohen98}), although age differences of $\sim$2\,Gyrs are still allowed within uncertainties.  
The colour distributions of GCs in NGC\,4365 and NGC\,5846 do not show the bimodality feature (\citealt{bs06}) as clearly, as other high mass galaxies. Instead, NGC\,4365 has been reported to have a colour distribution with a normal blue peak if compared to other galaxies but a broader red peak shifted towards bluer (V-I) colours (\citealt{larsen01}, \citealt{lbs05}). \cite{peng06} report bimodality for the $(g-z)$ colour distribution of this galaxy with a broad red peak. The colour distribution of NGC\,5846 is bimodal although the distribution of the central clusters shows a red peak with a hint of some additional intermediate colour GCs (\citealt{forbes97}). Adding the high fraction of intermediate-age GCs to this argument, suggests that the GC systems of these galaxies are not like other similar type ellipticals.

The apparent inconsistent results of NGC\,4365 and NGC\,5846 with no sign of intermediate-age stellar populations hosting a large fraction of intermediate-age GCs is intriguing.
GC studies and integrated light studies of the stellar population content of a galaxy should yield compatible conclusions.
With this in mind we have studied the ages of GC systems in 14 E/S0 galaxies through optical/NIR photometry.
We compare differentially a large number of GC systems with homogeneous data. The GC systems are also compared with SSP models.

\section{Observations and data}
The homogeneous survey of extragalactic GCs that we carried out with optical and NIR photometry was presented in \cite{paper1}.
In the same article the data reduction techniques were described in detail, here, we will only briefly mention them.
We obtained K ($K_S$, 2.2$\mu$m) band imaging with the LIRIS spectrographer and imager (\citealt{apulido03}, \citealt{manchado04}) at the William Herschel Telescope (WHT) from 2007 to 2009. The K imaging was combined with archival F475W ($\sim$ Sloan g) and F850LP ($\sim$ Sloan z) images from the Advanced Camera for Surveys (ACS) on board the Hubble Space Telescope (HST). 
Our sample is a sub-sample of the SAURON sample (\citealt{zeeuw02}) and comprises bright elliptical and lenticular galaxies ($M_B\,<\,19$) with $(m-M)\,< \,32$.
Each galaxy was imaged in the K-band for about 3.4 hours and reduced with LIRISDR and standard IRAF routines.    
The g and z images were reduced with the MULTIDRIZZLE task in the STSDAS package. 

The GCs were detected in the ACS images and transformed to the LIRIS coordinate system. 
Following this, aperture photometry and effective radii ($R_{\rm eff}$) were measured for the cluster candidates, with PHOT (\citealt{stetson87}) and ISHAPE (\citealt{larsen99}), respectively. 
A final sample of GCs was obtained applying the criteria:  $g<23$, $0.5<(g-z)<2.0$, $1<R_{\rm eff}({\rm pc})<15$.
Finally, a careful visual inspection was done, where obvious background galaxies still left in the sample were excluded. Sources too close to each other in the ACS images were flagged and if they appeared as one bigger source in the LIRIS frame they were also excluded.
We stress that any results for NGC\,4382 and NGC\,4473 should be taken with caution as their K-band observations were obtained in highly extincted conditions.

\section{Deriving ages of GCs}
In this section we attempt to derive ages of GC systems employing two different methods. In Sect 3.1 we directly compare the GC systems with SSP models.
In Sect 3.2 we look for age differences without making use of SSP models. The second method gives more promising results. 
\subsection{Simple stellar population models}
\subsubsection{Data-model comparison}

\begin{figure}
\begin{center}
\includegraphics[width=8cm]{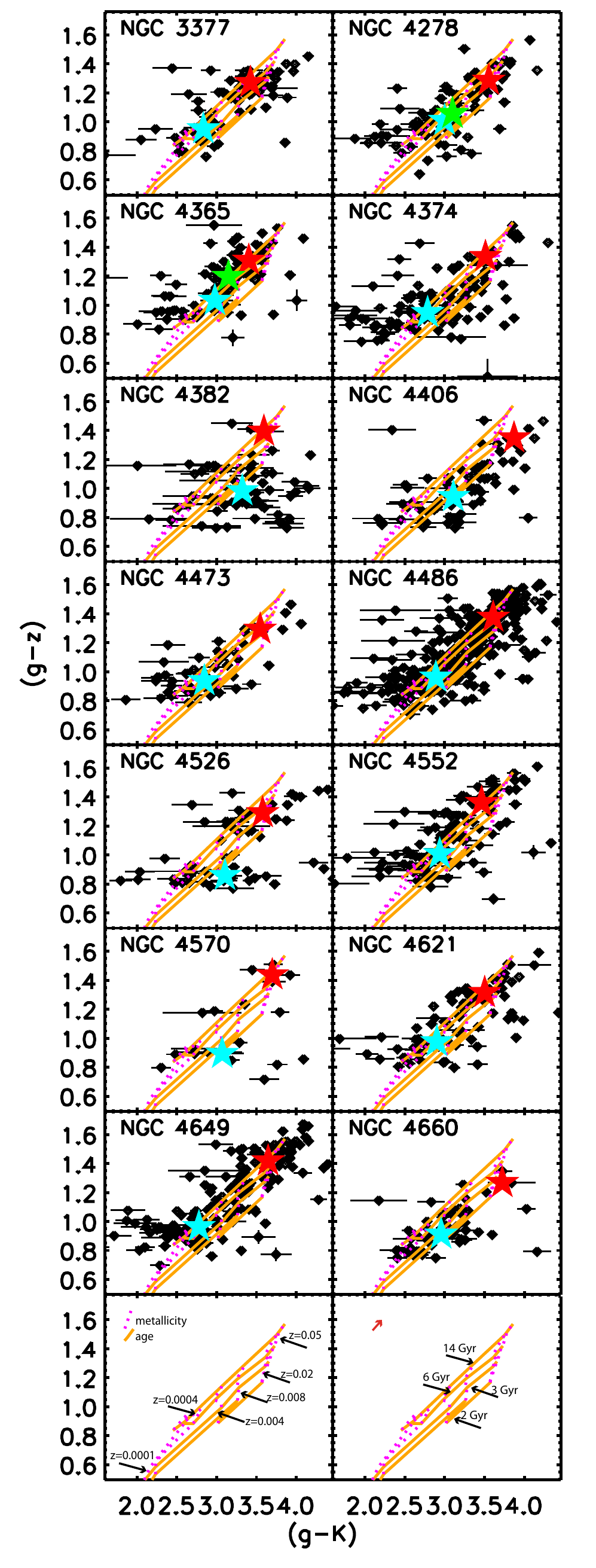} 
 \caption{2-colour diagrams for the GC systems of the 14 galaxies. $(g-z)$ \textit{vs.} $(g-K)$. Padova08 SSP models are overplotted with model sequences of constant age and metallicity according to the legend in the bottom panels. The ages increase from the lower right to the upper left. The metallicity tracks increase from the lower left to the upper right.
The blue and red larger symbols are the median of the blue and red sub-populations divided assuming a common division value of $(g-z)=1.187$. The green symbol in the panels corresponding to NGC\,4365 and NGC\,4278 indicate the median of the single populations as the $(g-z)$ distribution of these galaxies was considered to be more likely unimodal. The red arrow on the bottom most right panel indicates a typical reddening vector.}
\label{2colpadova}
\end{center}
\end{figure}

In Fig. \ref{2colpadova} we plot 2-colour diagrams, $(g-z)$ \textit{vs.} $(g-K)$ for the final sample of GCs, following the criteria outlined in Sect. 2.
Padova SSP models are shown in this plot, with model sequences of constant age ranging from the lower right to the upper left of each panel (2, 3, 6, 14\,Gyrs), see legend in the bottom panels. Lines of constant metallicity are also plotted. Metal poor clusters are expected to be located in the bottom left part of the diagram whereas metal rich ones should be located in the upper portion.
Padova SSPs are retrieved from the CMD 2.2 input form (http://stev.oapd.inaf.it/cmd), with \cite{marigo08} isochrones, and are from now on referred to as Padova08 SSPs. 
Compared with these models, the data appears to be consistent with the oldest track, 14\,Gyrs. Nevertheless, the 6\,Gyr intermediate-age track also seems to be a good match.
Excluding NGC\,4382 due to highly extincted data, qualitatively NGC\,4406 seems to host GCs with perhaps younger ages than the rest of the sample, while NGC\,4365 does not appear to differ significantly from the other galaxies in terms of the ages of its GCs.
Differences and similarities of the GC systems are quantified in Sects. 3.1.2 and  3.2.

\begin{figure*}
\begin{center}
\includegraphics[width=10cm]{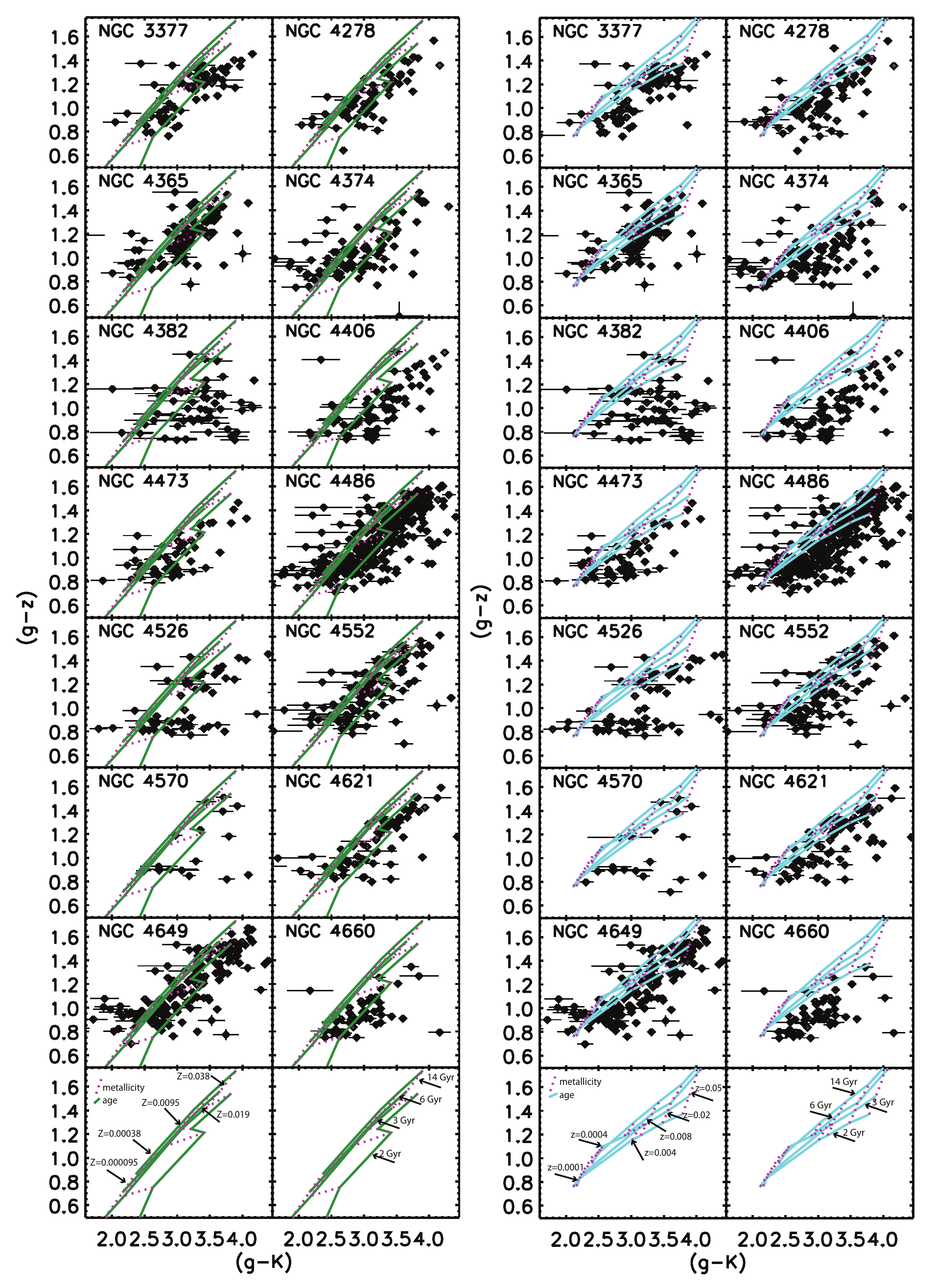} 
 \caption{2-colour diagrams for the GC systems of the 14 galaxies. $(g-K)$ \textit{vs.} $(g-z)$, as in Fig. \ref{2colpadova}: \cite{m05} with red horizontal branch at the two lowest metallicities \textit{(left panel)} and \cite{cb07} \textit{(right panel)}.}
\label{2col}
\end{center}
\end{figure*}

In Fig. \ref{2col} \cite{m05} and \cite{cb07} models are overplotted in the 2-colour diagrams, as in Fig. \ref{2colpadova}.
The comparison of the data with these models (different panels of Fig. \ref{2col}) would suggest younger ages than the comparison with the Padova08 models.
Therefore, concluding how old or young a GC is through direct comparison to SSP models does depend on the choice of the model.
Nevertheless, ages estimates through Padova08 models are closer to spectroscopically derived ages.
This is probably due to the new AGB and TP-AGB treatment of \cite{marigo08}. 
\cite{m05} uses a different prescription of the AGB and TP-AGB phases through the fuel consumption theorem. 
Charlot and Bruzual models use older versions of the Padova tracks (e.g. Padova 1999) which include older calibrations of the AGB phase and do not include a detailed prescription of the TP-AGB phase. 

\subsubsection{Blue and red peak ages}
In Fig. \ref{2colpadova} the median values of the blue and red populations illustrated with larger symbols are also shown. We have divided each GC system in a red (metal-rich) and a blue (metal-poor) sub-population.
We have chosen to use the median values as the observational scatter is colour dependant (with blue clusters having more spread) and the median is more robust against outliers. The value which indicates the division between blue and red clusters was determined by means of the KMM test (\citealt{abz94}). We fit two Gaussians to the $(g-z)$ distributions of the galaxies.
The value where the blue and red Gaussians intersect is the $(g-z)$ division value for the sub-populations. We use the mean of this value for the two most cluster-rich galaxies, NGC\,4486 and NGC\,4649, $(g-z)=1.187$. 
For NGC\,4365 and NGC\,4278 the $(g-z)$ distribution was found to be most likely unimodal. For this reason we also plot the median of the whole population with a different colour.
The median of the blue sub-populations have a trend of falling on relative younger tracks than the median of the red sub-populations. For example the value which corresponds to the median of the red sub-population of NGC\,4486 is located between the 6 and 14\,Gyr SSP while the blue one falls between the 3 and 6\,Gyr SSP. 

\begin{figure}
\begin{center}
\includegraphics[width=8cm]{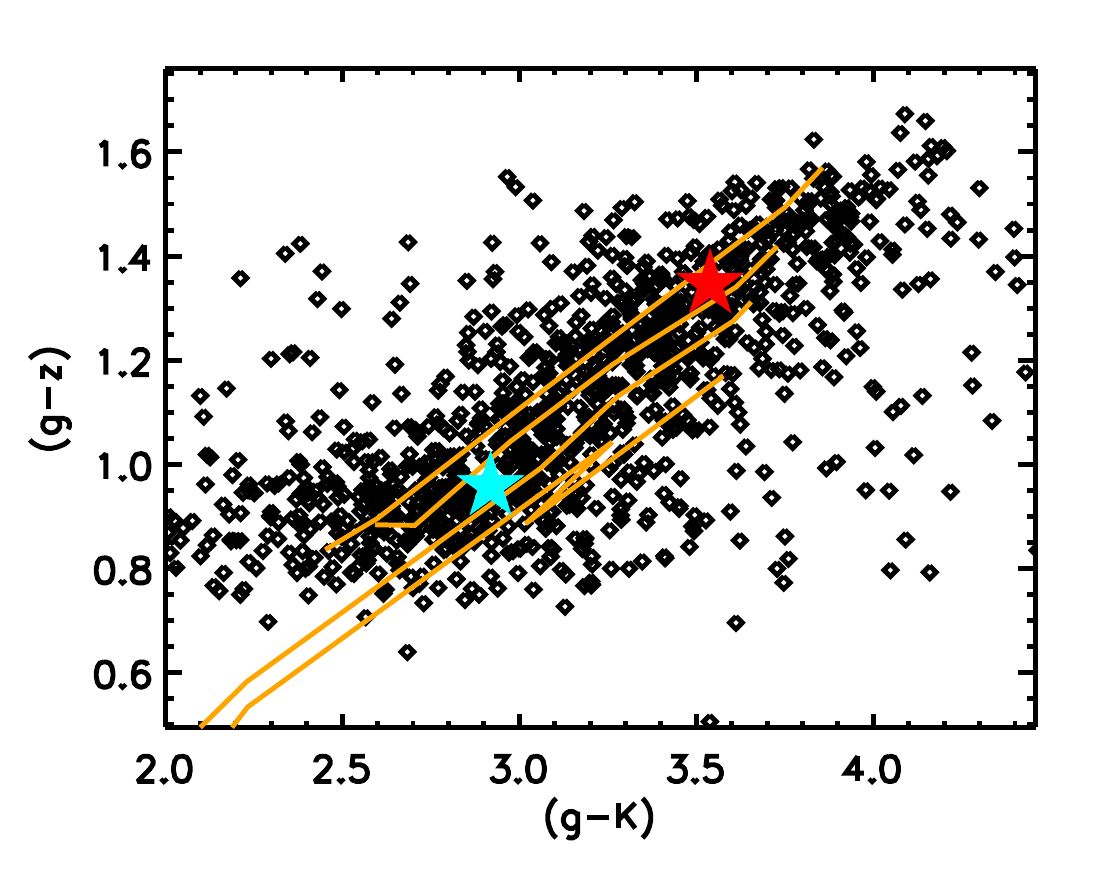}
\caption{The $(g-K)$ \textit{vs} $(g-z)$ 2-colour diagram for the GCs of 12 galaxies combined, excluding NGC\,4382 and NGC\,4473. The blue and red large symbols correspond to the median colour of the blue and red clusters respectively. The orange lines represent Padova08 SSPs of 2, 3, 6 and 14 Gyrs, like in Fig. \ref{2colpadova}.}
\label{14galaxies}
\end{center}
\end{figure}

In Fig. \ref{14galaxies} a $(g-K)$ \textit{vs} $(g-z)$ 2-colour diagram for the GCs of 12 galaxies combined, are shown excluding NGC\,4382 and NGC\,4473. The blue and red large symbols correspond to the median colour of the blue and red clusters respectively, assuming a fixed dividing colour at $(g-z)=1.187$.
This Fig. shows that the blue population is subject to larger scatter in the $(g-K)$ direction than the red population.
This is due to the fact that blue clusters are generally fainter in $K$, especially due to the selection in the $g$ band. This is also seen in Fig. \ref{2colpadova}.
The finding of on average younger blue GCs if compared to red ones in Figs. \ref{2colpadova} and \ref{14galaxies} is attributed to the direct comparison between data and SSP models. We stress that this should be interpreted with caution and verified with spectroscopy and/or with future SSPs.
One might think that objects with large $(g-k)$ colours and small error bars (see Fig. \ref{2colpadova}) would be making the blue population artificially young. Part of these objects could be background stars or foreground galaxies (see e.g. \citealt{puzia04}).
We note that a careful visual inspection was performed removing obvious background galaxies and blends (see \citealt{paper1}). There could still be some contaminants left, but only spectroscopic analysis of such objects would give more clear answers.

In order to quantify the ages that best match the GC systems we ran AnalySED (\citealt{anders04}) on the median magnitudes of the sub-populations, presented in Fig. \ref{2colpadova}. This code finds the best age and metallicity fits using model spectral energy distributions (SEDs). We used as input SEDs Padova08 SSPs integrated colours and kept the extinction fixed. 
For simplicity, Padova08 models starting at $1$\,Gyr were employed. 
At old ages, metallicity has a much stronger effect on the SEDs than does age, as seen in the 2-colour diagrams shown before.
Therefore, to get meaningful results it is important to have a grid in metallicity with sufficient resolution. For Padova08 SSPs one can download a grid of metallicities with a spacing that suits the user.

\clearpage
\begin{figure}
\begin{center}
\includegraphics[width=9cm]{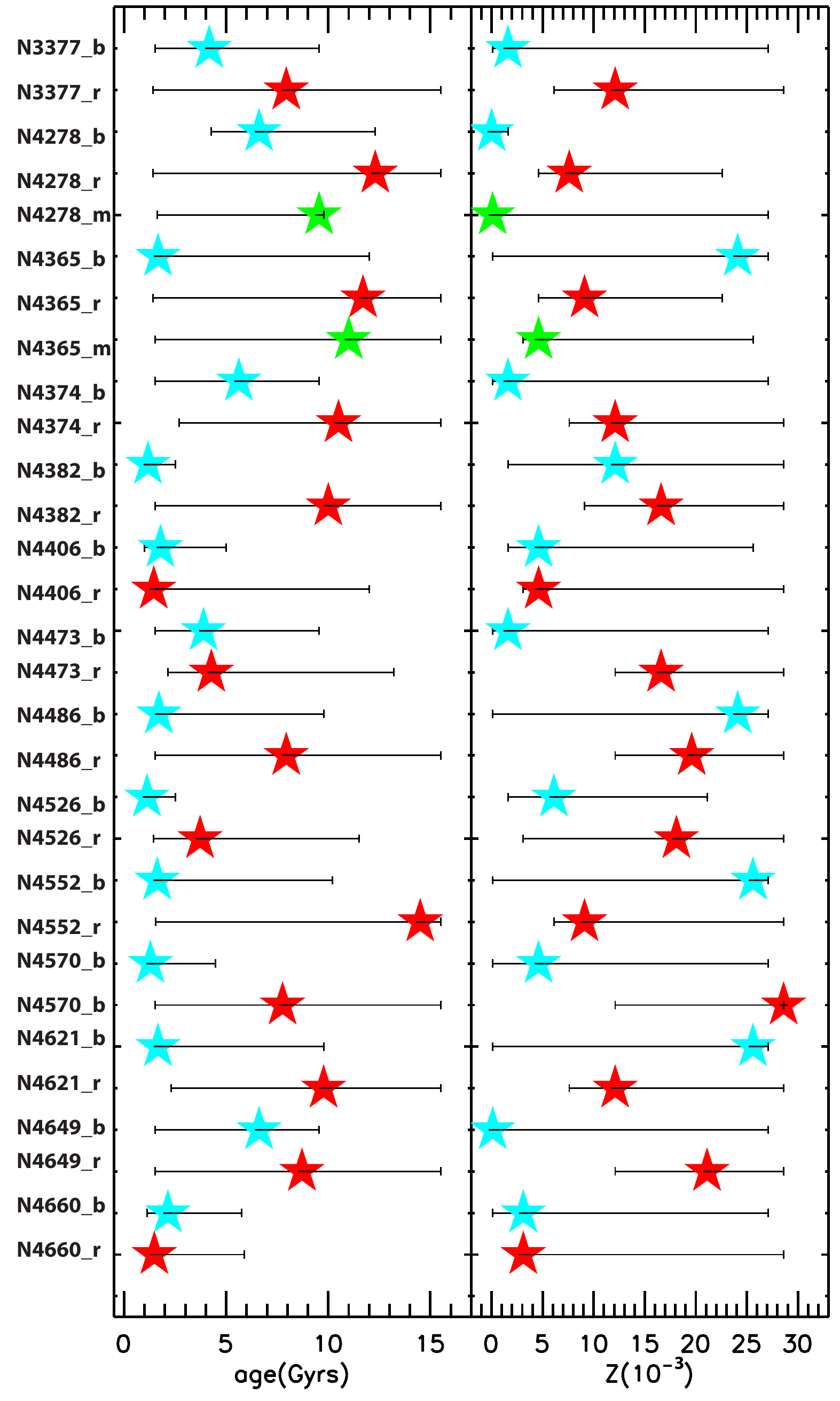}
\caption{The best age and metallicity fits returned by AnalySED with Padova08 SSPs for the median of the blue (metal poor) and red (metal rich) sub-populations indicated by blue and red symbols respectively. The median of a single population for NGC\,4365 and NGC\,4278 where KMM rejected the bimodal fit as opposed to the unimodal one is indicated by a green star.}
\label{AnalySED}
\end{center}
\end{figure} 

In Fig. \ref{AnalySED} the best age and metallicity fits returned by AnalySED are shown. The error bars represent the standard error on the mean. We have used a metallicity grid with a spacing of $\Delta\,$Z=0.0015. The ages derived by AnalySED vary between 1 and 14 Gyrs, \textit{i.e}, the data is roughly consistent with the whole range of models.
The upper limits in most cases however, appear to match our expectations (from spectroscopy, e.g. \citealt{strader05}, \citealt{cenarro07}) of the age of GCs in early type galaxies consistent with $\sim$10\,Gyrs. 
We strongly caution against interpreting the ages derived in the present section as real ages, they only serve to give an impression of what one gets from a literal comparison of the data with SSP models.
Mean ages are well below 10\,Gyrs, formally the mean blue is $\sim$3\,Gyrs while the mean red is $\sim$8\,Gyrs. 
Note that this is the same qualitative conclusion reached with Figs. \ref{2colpadova} and \ref{14galaxies}: the blue population is on average younger than the red one by using the Padova08 SSPs as a reference. 
The corresponding best metallicity fits are shown in the right panel of Fig. \ref{AnalySED}. The median blue best fit is Z=0.00046 ([Fe/H]$\sim-1.61$ dex) and the red is Z=0.00121([Fe/H]$\sim-1.19$ dex)\footnote{The approximation [Fe/H]$\sim$log(Z/Z$\sun$) and Z$\sun$=0.019 were used.}. 
The metal-poor peak value is comparable, or somewhat lower than what is found for typical GC systems of massive galaxies.
For example \cite{cohen98} finds the blue and red metallicity peak values for the NGC\,4486 GC system to be $-1.3$ and $-0.7$\,dex respectively, with a dispersion of $0.3$\,dex.
More recent measurements (\citealt{bs06}) agree on the blue peak value, $-1.3$ but the red peak is significantly more metal-rich, $-0.2$\,dex.

\subsection{Relative comparison without SSP models}
Until up to this point we have attempted to derive ages by direct data model comparison.
It is well documented in the literature that a direct comparison
of integrated GC colours with SSP models tends to yield ages that
are significantly younger than a Hubble time (e.g. \citealt{lbs05}). While this effect is less pronounced in the most recent Padova 
SSP models compared to earlier models, we still find it difficult to 
believe the very young ages assigned to the blue GCs in particular 
(Fig. \ref{AnalySED}).  It is worth noting that the spacing between isochrones
of different ages remains small even when including near-infrared
colours (Fig. \ref{2colpadova}), so even small errors in the model colours will have 
a large effect on the ages.  We expect that as the models continue
to be refined, these offsets will grow smaller. In the meantime,
it may be fruitful to explore other ways of determining age
differences between GC systems in different galaxies. We therefore
now turn to a purely differential comparison, independent of any
SSP models.

The median of the age distribution of the GC system of NGC\,4486 is reported to be 13\,Gyrs with a dispersion of 2\,Gyrs, through spectroscopy (\citealt{cohen98}). The GCs of NGC\,4649 are distributed in a similar way in the $(g-K)$ \textit{vs.} $(g-z)$ plane to those of NGC\,4486. Being these two GC systems supposedly old, and the ones with more clusters (167 and 301 respectively) we take them as the fiducial old systems. We fit linear relations to the data in the $(g-K)$ \textit{vs.} $(g-z)$ plane. These fits give more weight to the objects with smaller observational errors in both the abscissa and the ordinate.

\begin{figure}
\begin{center}
\includegraphics[width=7cm,angle=90]{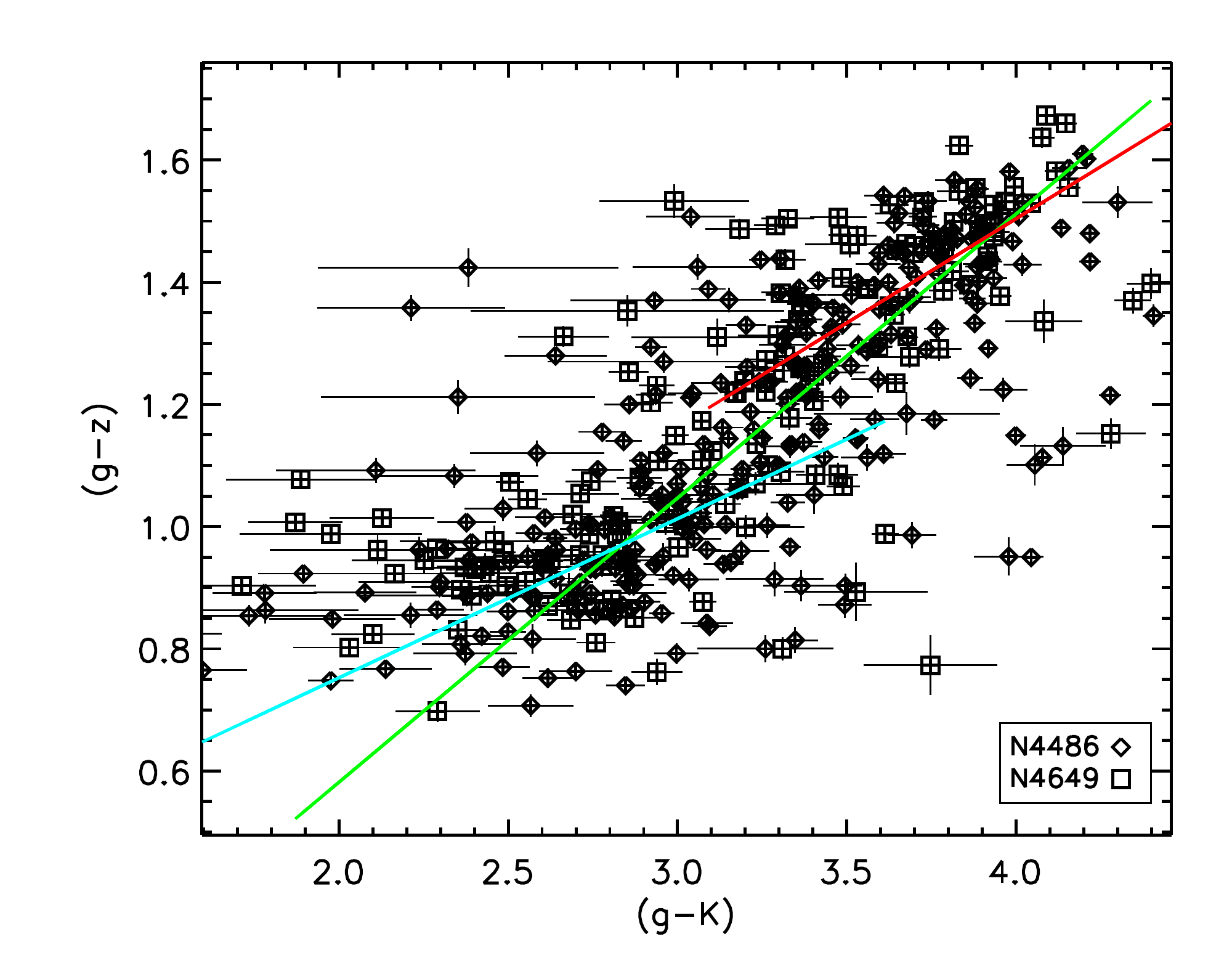}
\caption{$(g-K)$ vs. $(g-z)$ for the joint GC systems of NGC\,4486 and NGC\,4649. The weighted best fit for the whole sample, given by Eq. (1) is over plotted in green. The best fit lines for the blue and red sub-populations, given by (2) and (3) are shown in blue and red respectively.}
\label{linrel}
\end{center}
\end{figure} 

In Fig. \ref{linrel} this best fit relation given by Eq.\,(1) is shown. In the same plot, the best fit for the blue and red sub-populations are also shown. These, are given by Eqs. (2) and (3) respectively. 
Eq.\,(2) is valid for $(g-z)\,<\,$1.187 and Eq.\,(3) for  $(g-z)\,>\,$1.187. 
One possible reason for the apparent discontinuity of the $(g-z)$ \textit{vs.} $(g-K)$ relations of the blue and red GCs is that the horizontal branch morphology is expected to change abruptly at intermediate colours-metallicities, as discussed in Sect. 4.2. A separation into blue and red best fit lines is taking this into account. 

\begin{equation}
(g-z)=0.465*(g-K)-0.349
\end{equation}

\begin{equation}
(g-z)_{b}=0.260*(g-K)_{b}+0.232
\end{equation}

\begin{equation}
(g-z)_{r}=0.340*(g-K)_{r}+0.140
\end{equation}

\begin{figure}
\begin{center}
\includegraphics[width=8cm]{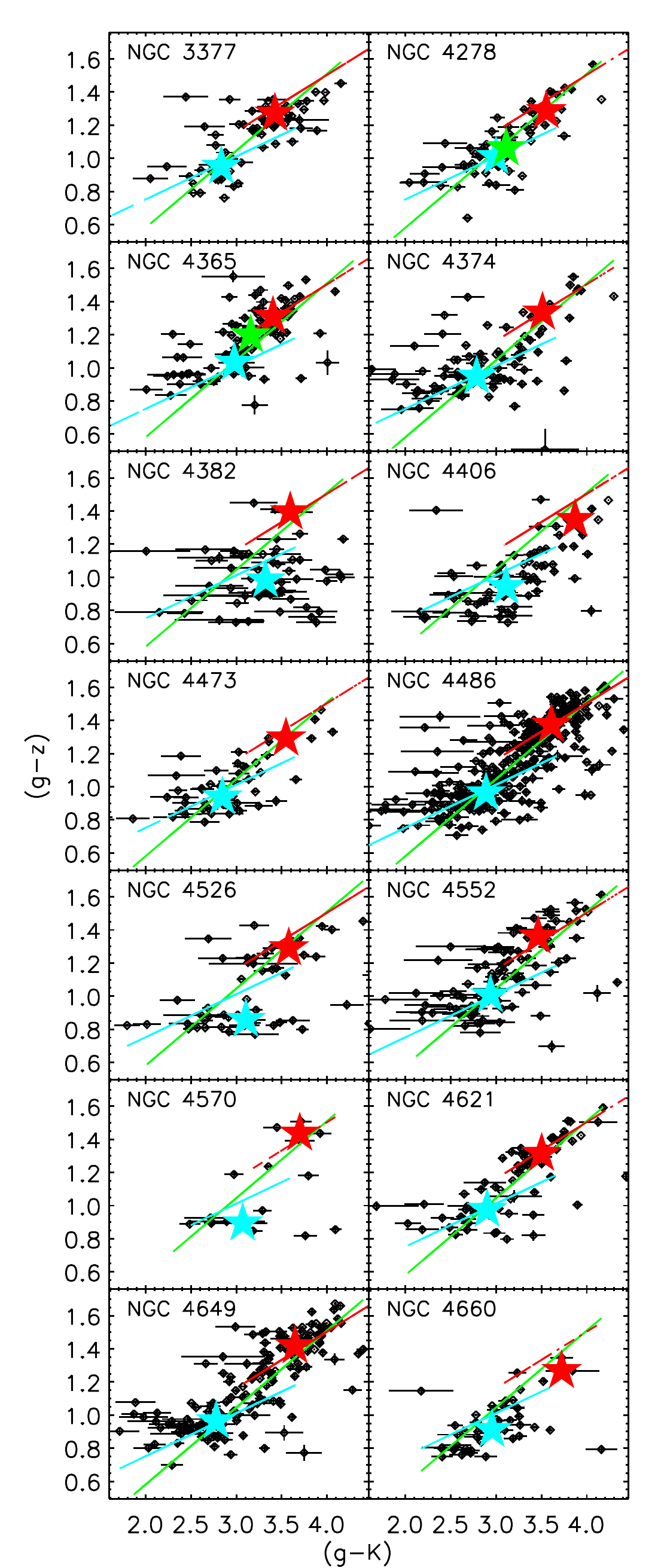}
\caption{Same as Fig \ref{2colpadova} but with the Padova08 SSPs replaced with the best fit relations given by Eqs. (1), (2) and (3) shown in Fig \ref{linrel} over plotted.}
\label{bestfitm60m87_allgalaxies_bluered}
\end{center}
\end{figure}

In Fig. \ref{bestfitm60m87_allgalaxies_bluered} the best fit lines (1), (2) and (3) in the $(g-K)$ vs. $(g-z)$ plane for the joint GC systems of NGC\,4486 and NGC\,4649 are over plotted for the different galaxies.
Also plotted are the median colours for the blue and red sub-populations of the respective galaxies.  
This plot shows qualitatively how much and in what direction the mean of the sub-populations deviates from the lines in Fig. \ref{linrel}.
For instance, the median of the blue and red populations of some GC systems, such as NGC\,4406, deviate significantly to the right of the green line. 

\begin{figure}
\begin{center}
\includegraphics[width=8cm]{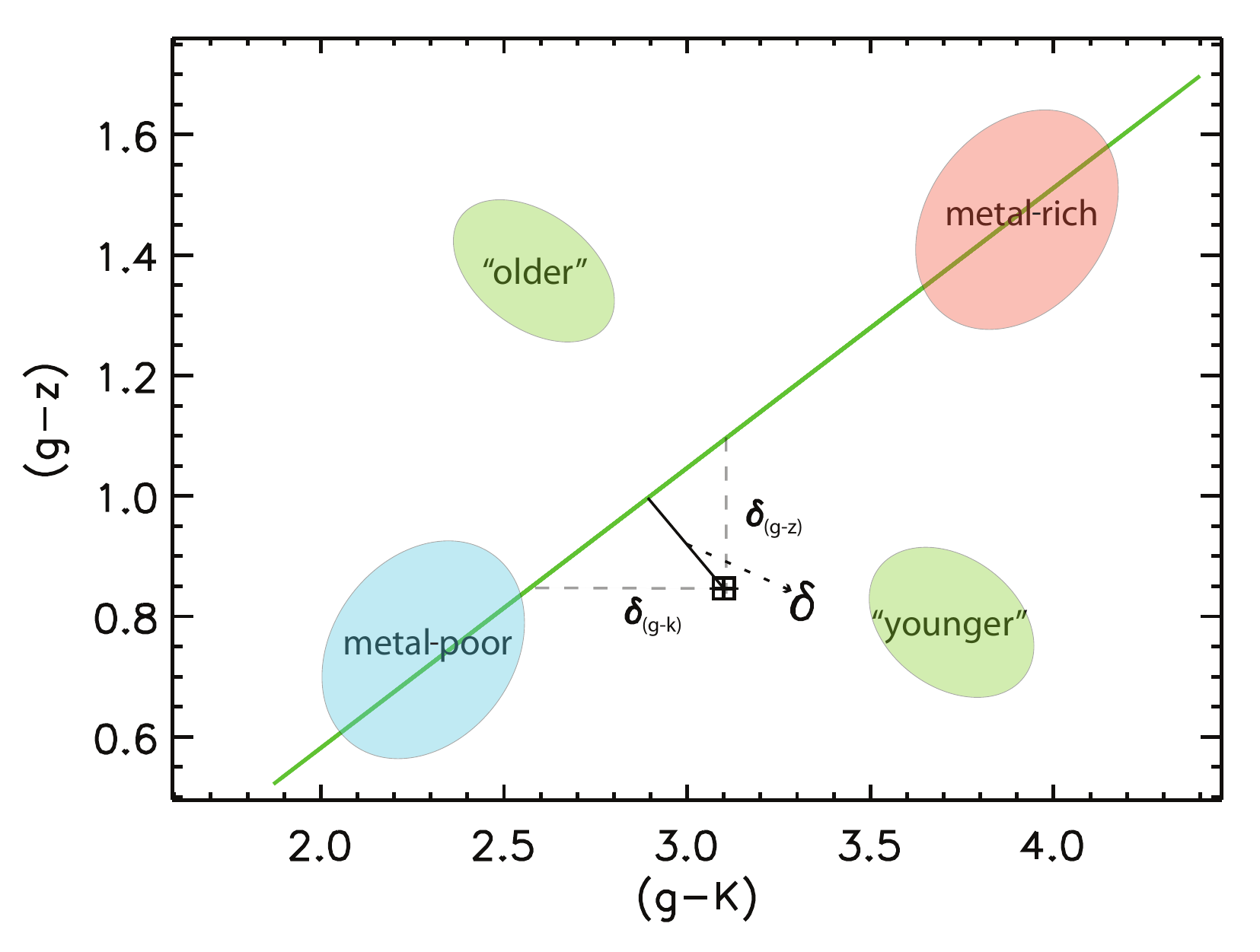}
\caption{A cartoon illustrating the definition of $\delta$ in the $(g-K)$ vs. $(g-z)$ diagram for one cluster (see text for details) and the location of the "younger" and "older" objects relative to the green line. The location of metal-poor and metal-rich objects is also shown.}
\label{cartoon}
\end{center}
\end{figure} 

In the following,
we introduce a new method to derive \textit{relative} ages of GC systems with the purpose of quantifying the magnitude and direction of the deviation of a GC from Eqs. (1) (2) and (3).
In Sect. 3.1 we showed that relative to the SSP models, a younger cluster falls on the lower right of the $(g-K)$ vs. $(g-z)$ plane while an older cluster falls on the upper left of this diagram.
By considering the green line as an SSP, with metallicities increasing from the lower left to the upper right, one can make statements about relative ages of GCs with respect to this line.
We define $\delta$, $\delta_{\rm blue}$ and $\delta_{\rm red}$ as the projected distances of a GC from relations (1), (2) and (3) respectively. These, are given by:

\begin{equation}
\delta=\frac{0.465*(g-K)-(g-z) -0.349}{\sqrt{0.465^2+1}}
\end{equation}

\begin{equation}
\delta_{\rm blue}=\frac{0.260*(g-K)-(g-z) +0.230}{\sqrt{0.260^2+1}}
\end{equation}

\begin{equation}
\delta_{\rm red}=\frac{0.341*(g-K)-(g-z) +0.140}{\sqrt{0.341^2+1}}
\end{equation}

Fig. \ref{cartoon} shows a cartoon illustrating $\delta$ for a cluster in the $(g-K)$ vs. $(g-z)$ plane. In this figure the location of younger and older clusters relative to this line as well as the location of the metal rich and metal poor clusters are plotted. The values of $\delta$ are age tracers of GCs in this plane. 
An object falling to the right of the green line, will have a positive $\delta$ value. A positive $\delta$ value means that this object has an age that is younger than the age that the green line traces ($\delta=0$).
In this sense a GC system whose mean difference is further away to the right of the $\delta=0$ value in the distributions are the systems with the best chances of hosting younger GCs.

\begin{figure}
\begin{center}
\includegraphics[width=8cm]{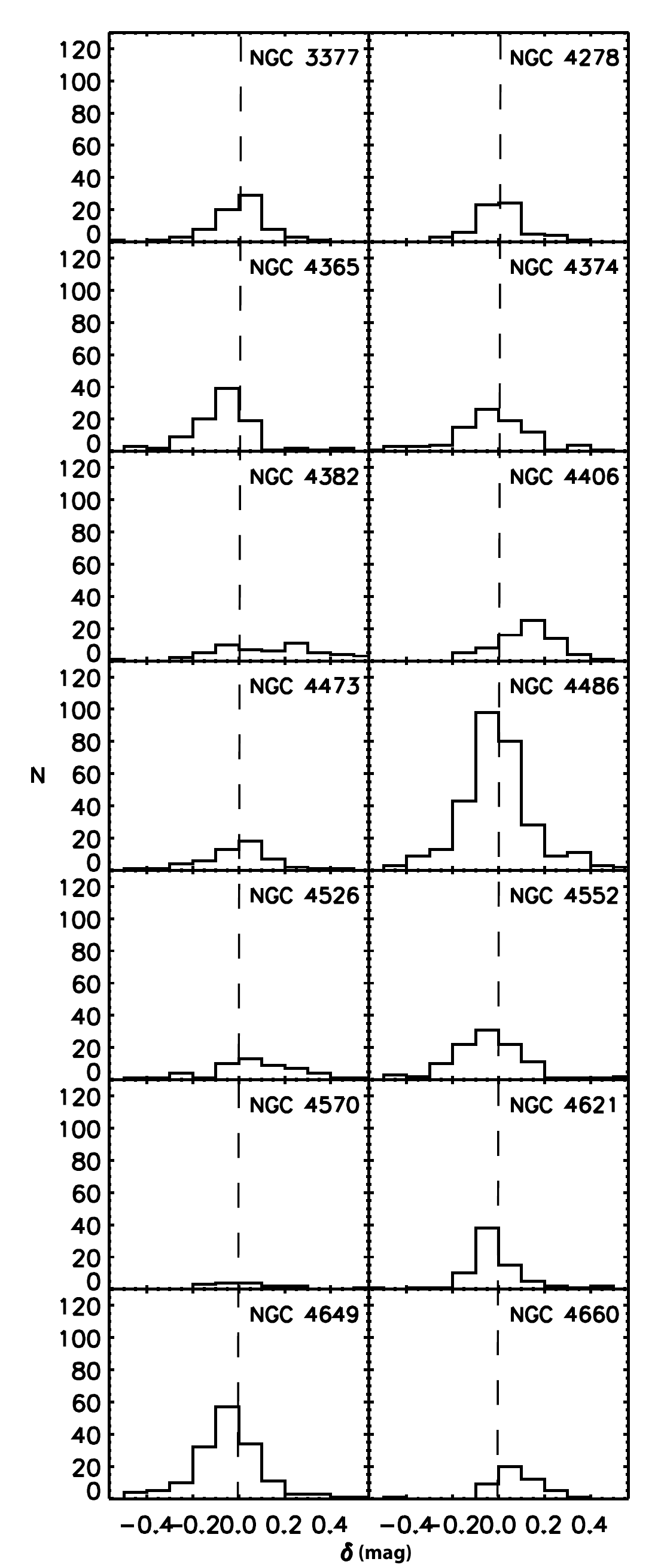}
\caption{The $\delta$ distribution: the difference between the GCs and the best fit line for the whole sample given by Eq. (1). The dashed line indicates the value $\delta=0$.}
\label{diffbestfitm60m87_allgalaxies}
\end{center}
\end{figure}

\begin{figure}
\begin{center}
\includegraphics[width=8cm]{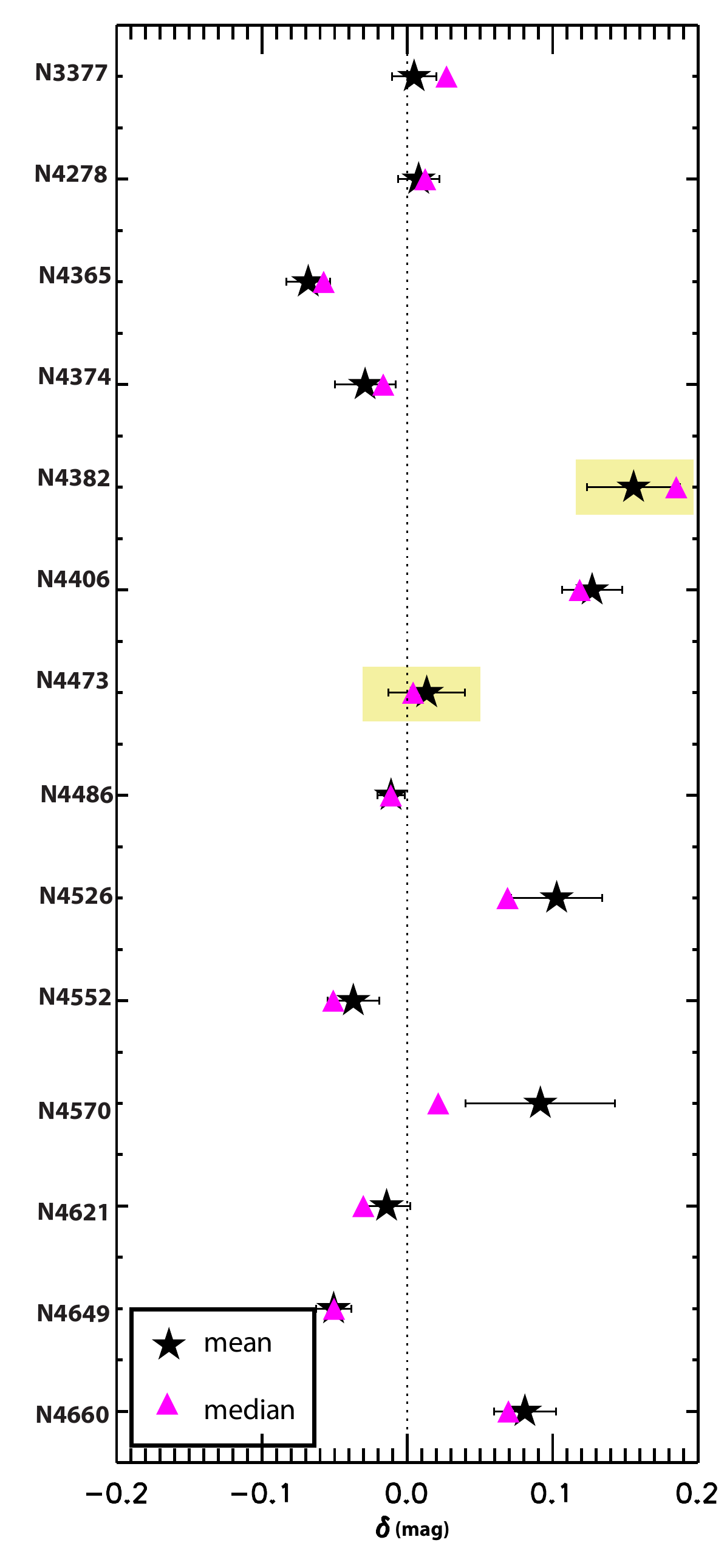}
\caption{The mean and median of the $\delta$ distribution for the GCs of different galaxies, shown in Fig. \ref{diffbestfitm60m87_allgalaxies}. The error bars centred in the mean indicate the standard error on the mean of the quantity $\delta$. The dotted line stands for $\delta=0$. NGC\,4382 and NGC\,4473 are marked with yellow rectangles because their K-band data were obtained in highly extincted conditions.}
\label{bestfitdiffout}
\end{center}
\end{figure}

In Fig. \ref{diffbestfitm60m87_allgalaxies} the histograms for the $\delta$ values of the different galaxies are shown.
By comparing the different $\delta$ distributions one can check if there are age differences between the different distributions.
It is readily seen that not all GC systems seem to be consistent with a $\delta$=0.
For example, note that the peak of the NGC\,4406 $\delta$ distribution is $\sim$0.2 whereas the peak of the NGC\,4649 is $\sim-$0.05.
By taking the mean value of the distributions of $\delta$ it is possible to see how on average the different GC systems are younger (or older) from the joint GC system of NGC\.4486 and NGC\,4649.
In Fig. \ref{bestfitdiffout} a diagram shows the mean (and median) of $\delta$ with the standard error on the mean for each GC system. The usual interpretation of any offset from $\delta$=0  would be that there are (mean) age shifts. The larger shifts occur for $\delta>$0, i.e to younger ages. Therefore the GC system of NGC\,4406 is on average younger than the GC system of NGC\,4649 and NGC\,4486.
The best cases for hosting younger GCs are NGC\,4382, NGC\,4406, NGC\,4526, NGC\,4570 and NGC\.4660. All of them are S0s, except for NGC\,4660 which is an E5, following classification from NED. However, this galaxy is also quoted to be an E3/S0 by \cite{ferrarese06} who report boxy isophotes and a faint blue structure of about 2.5\arcmin\,with two spiral arms within an ACS pointing. 

\begin{figure}
\begin{center}
\includegraphics[width=8cm]{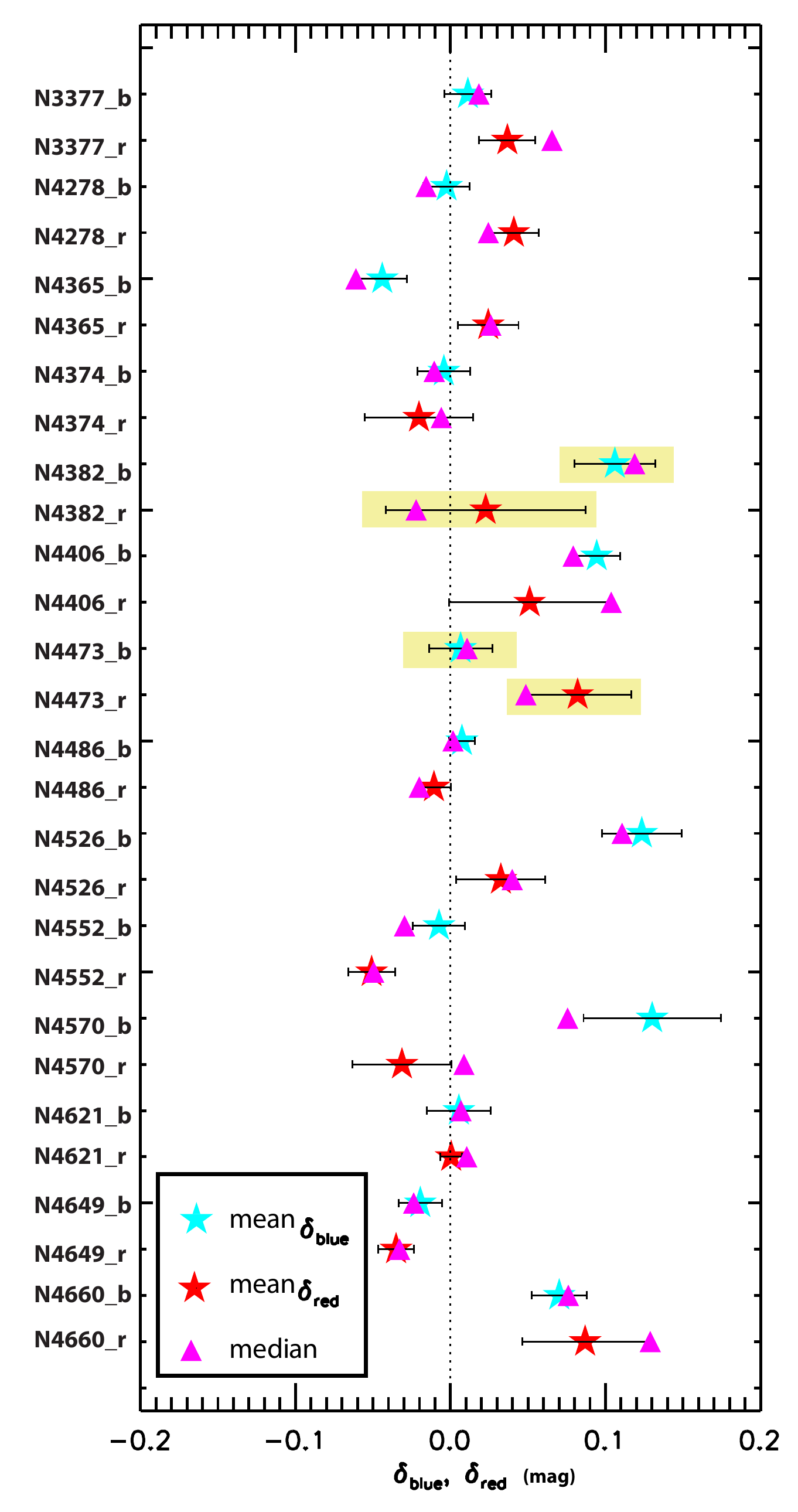}
\caption{Similar to Fig \ref{bestfitdiffout} but for the mean and median of the $\delta_{\rm blue}$ and $\delta_{\rm red}$ distributions for the GCs of different galaxies. The error bars centred in the mean indicate the standard error on the mean of the quantities. The dotted line stands for the value $\delta_{\rm blue}$ and $\delta_{\rm red}=0$. NGC\,4382 and NGC\,4473 are marked with yellow rectangles because their K-band data were obtained in highly extincted conditions.}
\label{bestfitdiffout_bluered}
\end{center}
\end{figure}

By performing the $\delta$ analysis for the different GC systems separate for blue and red clusters one can check which sub-population has a greater $\delta$ shift.
The means (and medians) of the $\delta_{\rm blue}$ and $\delta_{\rm red}$ distributions are plotted in Fig. \ref{bestfitdiffout_bluered}. 
The values of $\delta_{\rm blue}$ and $\delta_{\rm red}$ are based on the colour ranges where Eqs. (2) and (3) are valid.
The blue populations of NGC\,4570, NGC\,4526, NGC\,4406 and NGC\,4382 have the greatest shift, $\delta_{\rm blue} \ga 0.1$.
The red populations on the other hand, have an apparent smaller shift.
Moreover, the $\delta_{\rm blue}$ offsets from $\delta_{\rm blue}=0$ in Fig. \ref{bestfitdiffout_bluered} are very similar to the $\delta$ offsets from $\delta=0$ in Fig. \ref{bestfitdiffout}.

It is important to check how much scatter around the vertical line of $\delta$, $\delta_{\rm blue}$ and $\delta_{\rm red}=0$ might be attributed to zero point, photometric and aperture correction uncertainties. 
The accuracy of the LIRIS K-band photometric calibration is estimated to be $\sim0.03-0.05$\,mag, see Sect. 3.2 of \cite{paper1}.
There are cases where the mean $\delta$ reaches values as large as three times the values of the uncertainty on the photometric calibration.
Therefore only part of the scatter on Figs. \ref{bestfitdiffout} and \ref{bestfitdiffout_bluered} could be attributed to uncertainties in the K-band photometric calibration. 

\begin{figure}
\begin{center}
\includegraphics[width=9.5cm]{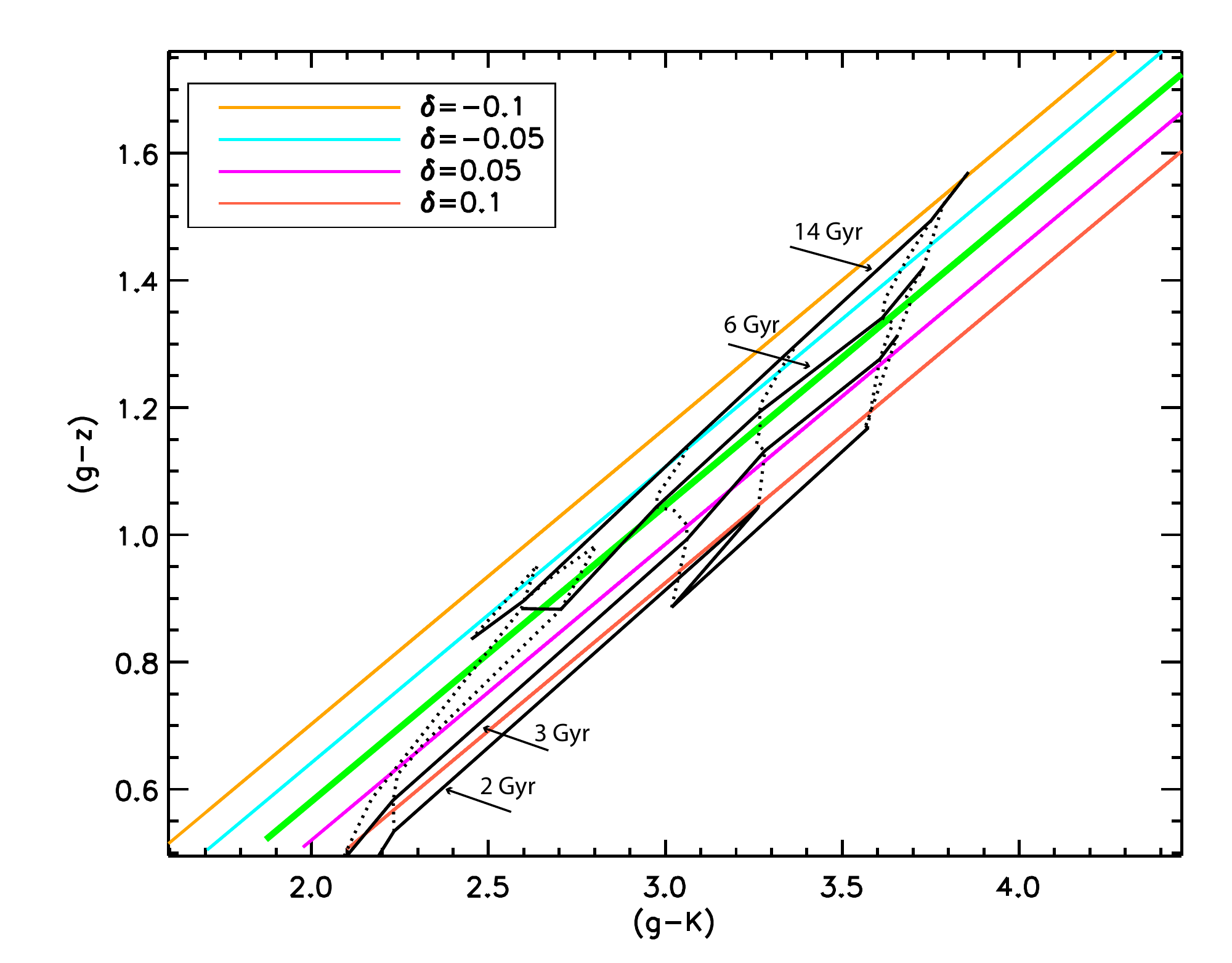}
\caption{$(g-K)$ vs. $(g-z)$ diagrams showing lines of constant $\delta$, according to the legend.
Padova08 SSP models are over plotted as in Fig \ref{2colpadova}. The green line is the best fit line for the joint GC systems of NGC\,4486 and NGC\,4649, given by relation (1).}
\label{delta_quantitatively}
\end{center}
\end{figure}

While measurements of the mean $\delta$ parameter in different galaxies
reveal some interesting differences, we have not yet discussed how to
interpret this in terms of actual age differences. To this end, we
are guided by a comparison with SSP models.
In Fig. \ref {delta_quantitatively} $(g-K)$ vs. $(g-z)$ 2-colour diagrams are shown with lines of constant $\delta$. Padova08 SSP models are also overplotted, as in Fig. \ref{2colpadova}.
Note that the best fit line (1), is roughly consistent with the 6\,Gyr track.
Moreover the line of $\delta\sim0.1$ is located between the 2 and 3 Gyr old tracks.
This means that an offset of $\delta$ $\sim$0.1 is roughly consistent with an upper limit of $\Delta$age$\sim$4\,Gyrs.
$\delta$ values consistent with $\sim$0.05 fall between the 3 and 6 Gyr old tracks, giving a maximum $\Delta$age$\sim$3\,Gyrs. 
If we instead shift the model grid such that the 14 Gyr SSP model
coincides with the $\delta=0$ line, an offset of $\delta=+0.1$
would correspond to an age of about 6 Gyr, or a $\Delta$age of 8 Gyrs.
It is thus clear that a translation of our $\delta$ parameter
to an actual age difference is quite uncertain.
We stress that care must be taken on the interpretation of the $\delta$-parameter as a pure proxy for age, due to the age-metallicity degeneracy. For instance, 
the SSP tracks in the 2-colour diagram do not run parallel to the lines of constant $\delta_{blue}$ and $\delta_{red}$ and the comparison between the two is
metallicity dependant.
Therefore it is difficult to even roughly estimate what $\delta_{\rm blue}$ and $\delta_{\rm red}$ mean in terms of $\Delta$age.

One might ask whether differences in the $\delta$ values could be related to metallicity rather than age effects due to the known peak GC metallicity \textit{vs.} galaxy luminosity relation (eg. \citealt{bs06}). This would happen because the fiducial best-fit lines Egs. (1), (2) and (3) are defined on two of the highest luminosity galaxies in our sample. We argue this does not appear to be the case. The reason is that the peak GC metallicity \textit{vs.} galaxy luminosity relation does not translate into a peak GC metallicity - morphology relation. For example, the S0 NGC\,4406 which has a higher delta value has a $M_B$ value comparable to that of galaxies with low delta values, eg. NGC\,4365. Another example is NGC\,4382 which has a high delta value and an $M_B$ value consistent with that of NGC\,4649. Although the case for NGC\,4382 is to be taken with caution due to the worse quality K-band data. 
Moreover, we have calculated the $\delta_{\rm blue}$ for Òtwo clustersÓ with the same age (10\,Gyrs) and with host galaxy luminosities, $M_B$=-21($Z\sim-0.0008$) and $M_B$=-19 GC ($Z\sim-0.0006$), following Fig. 13 of \cite{peng06}. This was done in order to test for the worst case scenario, the influence of the peak GC metallicity \textit{vs.} galaxy luminosity relation on the measured $\delta$ values. By making use of Padova08 SSPs we find that the difference between these $\delta_{\rm blue}$ values is 0.0064, i.e., much smaller than the actual delta values. Therefore, the peak GC metallicity \textit{vs.} galaxy luminosity relation has no significant impact on the $\delta$ values over the luminosity range relevant for our sample.

\subsection{GC ages correlated with host galaxy morphological type}
\begin{figure}
\begin{center}
\includegraphics[width=3.0in]{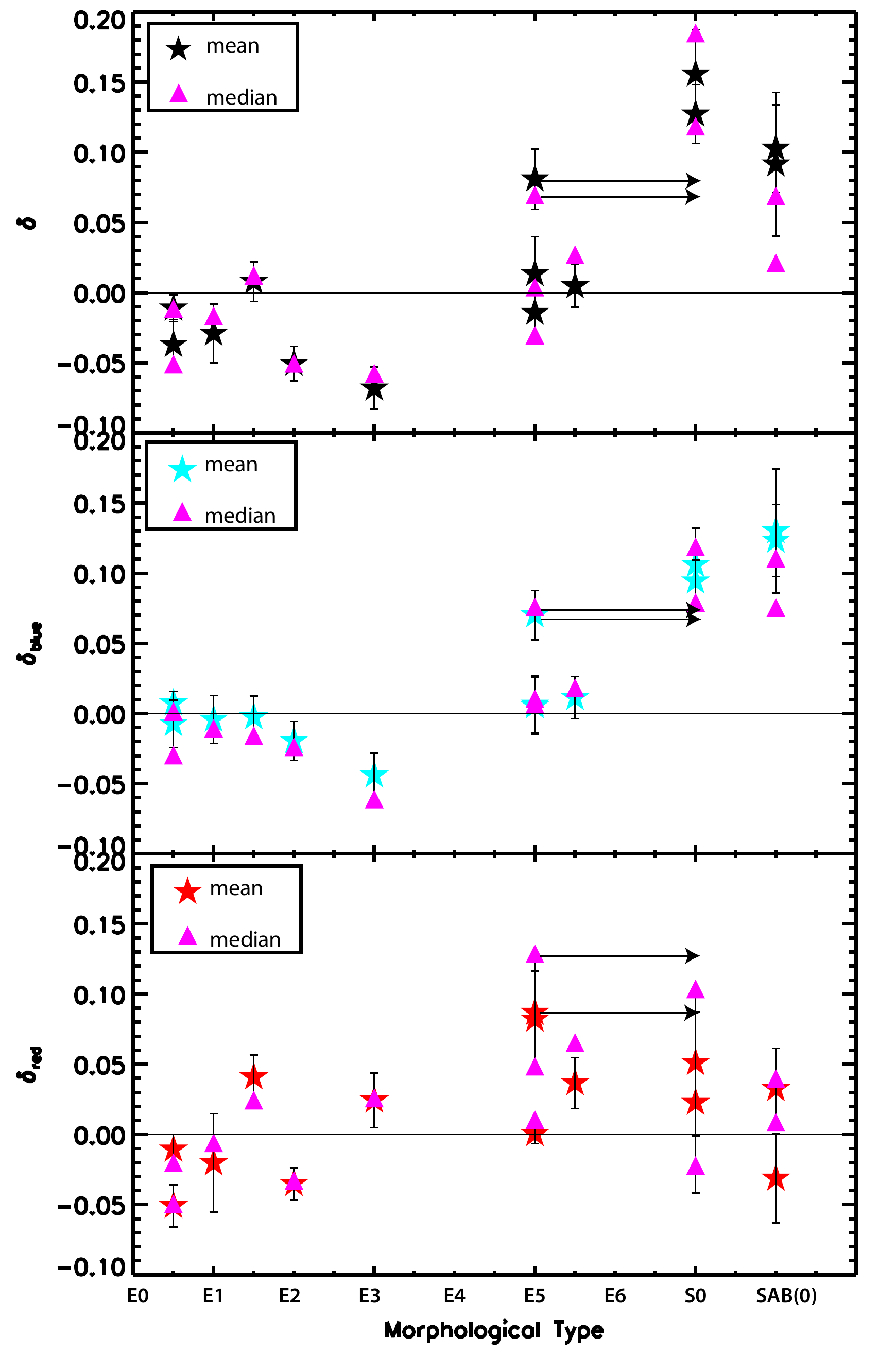} 
\caption{The mean and median of  $\delta$, $\delta_{\rm blue}$ and $\delta_{\rm red}$ as a function of morphological type in the different panels. The solid line indicates the value where the mean of the GCs would not differ from the relations (1), (2) and (3). The arrows indicate the location of the values for NGC\,4660 when classified as S0 instead of E5.}
\label{morph_bluered_all}
\end{center}
\end{figure}

In Fig. \ref{morph_bluered_all} the mean (and median) of $\delta$, $\delta_{\rm blue}$ and $\delta_{\rm red}$ are plotted against the morphological type of the galaxy.
The morphological classification of the galaxies is given in Table 1. It is readily seen that the mean of $\delta$ correlates with galaxy morphology.
The $\delta$ (and $\delta_{\rm blue}$, $\delta_{\rm red}$) distributions of the galaxies of later Hubble type (E5, S0's, SAB(0)'s) have on average larger positive offsets from $\delta=0$ (and $\delta_{\rm blue}=0$, $\delta_{\rm red}=0$) if compared to that of galaxies of earlier Hubble type (E0's). These offsets are interpreted as being due to on average younger GC ages.
If the E5 type galaxy NGC\,4660 were to be classified as an S0, this correlation would be even stronger.
To quantify the strength of this correlation we make use of the Spearman's rank correlation coefficient ($\rho$) which is a non-parametric measure of statistical dependence between two variables.
If one of the variables is a perfect monotone of the other the quantity $\rho$ will be equal to +1 for correlation and -1 for anticorrelation.
Considering NGC\,4660 as an E5(S0) in the relation of Fig. \ref{morph_bluered_all}, the values of $\rho$ are: 0.75(0.76) for the mean and 0.69(0.73) for the median. 
For $\delta_{\rm blue}$, $\rho$= 0.79(0.81), 0.81(0.83) and for $\delta_{\rm red}$, $\rho$=0.35(0.41), 0.47(0.53) for the mean and median respectively. 
If one removes NGC\,4382 and NGC\,4473, due to highly extincted K-band photometry, the $\rho$ values become 0.67, 0.63. For $\delta_{\rm blue}$ the $\rho$ values are 0.74, 0.76 and for $\delta_{\rm red}$ they are 0.41 and 0.65, for the mean and median respectively.
The values of $\rho$ for $\delta$ (and $\delta_{\rm blue}$) are close enough to 1 for the correlations between $\delta$ (and $\delta_{\rm blue}$) and galaxy morphology to be significant.

From Fig. \ref{delta_quantitatively} we have that an offset of $\delta=0.1$ is roughly consistent with a $\Delta$age of $\sim$\,4-8\,Gyrs.
If one considers the median age of $\sim$13\,Gyrs found by \cite{cohen98} for the GC system of NGC\,4486, an E0, the average ages of S0 GC systems are extrapolated to be $\sim$5-9\,Gyrs.
We stress that this is a very rough estimate, given the zero-point problem of the models.

\begin{table}
\centering
\begin{tabular}{ccc}

\hline
Galaxy & Morphological& Morphological\\
               &        Type            & Type (plots)\\
      (2)         &        (2)           & (3)\\
\hline
NGC 3377 & E5/E6 &E5/E6 \\
NGC 4278 & E1/E2&E1/E2\\
NGC 4365 & E3&E3\\
NGC 4374 & E1&E1\\
NGC 4382 & S0$^{+}(s)pec$&SAB(0)\\
NGC 4406 & S0&S0\\
NGC 4473 &E5&E5\\
NGC 4486 &E0/E1$^{+}pec$&E0/E1\\
NGC 4526 & SAB0$^{0}(s)$&SAB(0)\\
NGC 4552 &E0/E1&E0/E1\\
NGC 4570 &S0 sp&S0\\
NGC 4621 &E5&E5\\
NGC 4649 &E2&E2\\
NGC 4660 &E&E5\\
\hline
\end{tabular}
\caption{Galaxy (1), morphological type from \citealt{emsellem07} and NED (2) and morphological type used for Figs. \ref{morph_bluered_all} and \ref{morph_scatter}.}
\label{morphtab}
\end{table}

\section{Investigating effects other than age and metallicity in the 2-colour diagrams}
\begin{figure}
\begin{center}
\includegraphics[width=8cm]{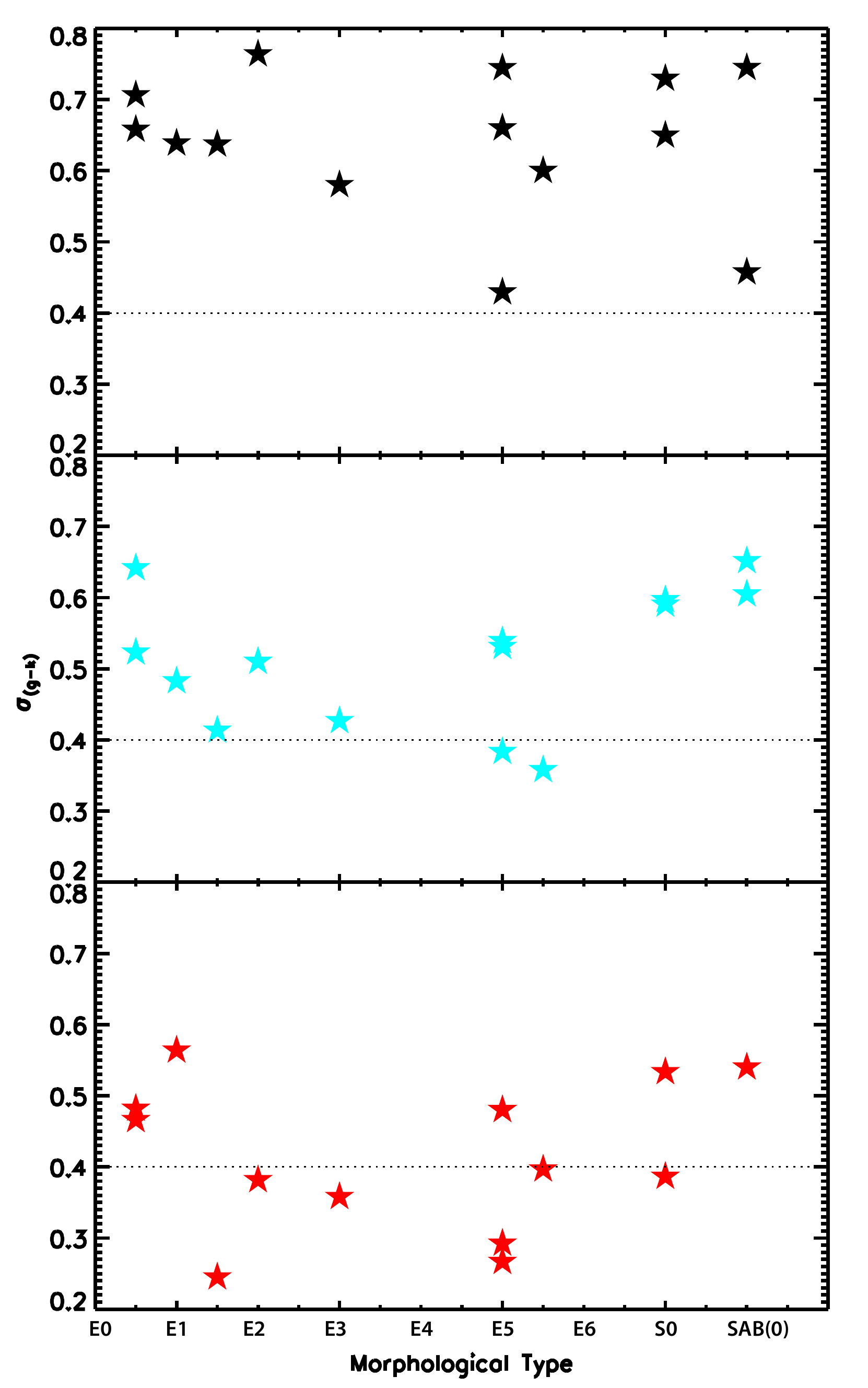} 
\caption{The scatter represented by the standard deviation ($\sigma$) for $\delta$, $\delta_{\rm blue}$ and $\delta_{\rm red}$ in the $(g-K)$ direction as a function of morphological type. The dotted lines indicate the upper limit to the maximum scatter from observational errors.}
\label{morph_scatter}
\end{center}
\end{figure}

In the top panel of Fig. \ref{morph_scatter} the standard deviation ($\sigma$) of the clusters around relation (1) along $(g-K)$ in the $(g-K)$ \textit{vs.} $(g-z)$ plane is shown as a function of morphological type. 
Because the photometric errors are dominated by the K-band
photometry, the scatter in (g-K) will be very nearly proportional
to the scatter in $\delta$ if photometric errors are the only
cause of offsets from the line defining the $\delta$ parameter.
It is immediately seen that the scatter varies among different systems and that in most cases it is larger than the 1$\sigma$ of $\sim\pm0.4$ permitted (\citealt{paper1}) from observational uncertainties. 
Analogously the scatter of $\delta_{\rm blue}$ and $\delta_{\rm red}$ along $(g-K)$ as function of galaxy morphology are shown in the middle and in the bottom panels respectively. It is readily seen that the scatter is greater for $\delta$ than for $\delta_{\rm blue}$ and $\delta_{\rm red}$ separately. 
For $\delta$, the mean $\sigma$ is $0.643\pm0.007$. The error quoted is the standard error on the mean. For $\delta_{\rm blue}$ the mean is $0.518\pm0.013$ and for $\delta_{\rm red}$ $0.403\pm0.015$.
The scatter becomes smaller when looking at $\delta_{\rm blue}$ and $\delta_{\rm red}$ separately. In fact, it seems almost consistent with the observational errors. This suggests that a linear fit in the 2-colour $(g-K)-(g-z)$ space is an over approximation. By making the division in two linear regimes, it seems that almost all the scatter can be explained by observational errors.

One might think that the relation shown in Fig. \ref{morph_bluered_all} might be an artefact of our sky subtraction method through ellipse model fits (see \citealt{paper1}). Even though the direction the morphology becomes more complicated is the same direction the ellipse model fits became more difficult, Fig. \ref{morph_scatter} shows that there is no relation between the scatter in $\delta$  ($\delta_{\rm blue}$ and $\delta_{\rm red}$) and the host galaxy morphological type. We therefore conclude that the observed trend of increasing
$\delta$ with Hubble type is real.

\begin{figure}
\begin{center}
\includegraphics[width=8cm]{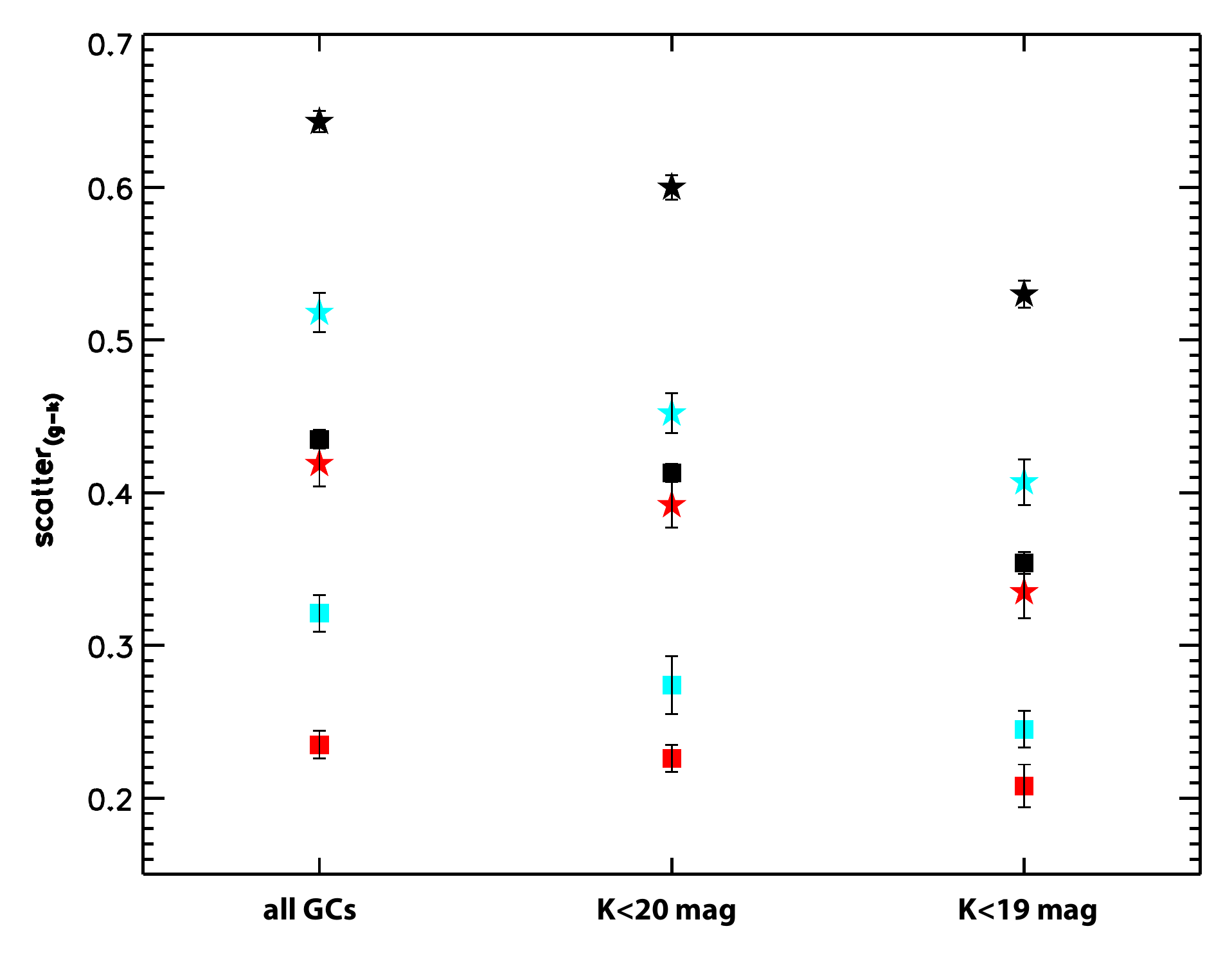} 
\caption{The mean $\sigma$ (stars) and the mean \textit{mad} (squares) for the measurements in Fig. \ref{morph_scatter} for $\delta$ (black symbols), $\delta_{\rm blue}$ (blue symbols) and $\delta_{red}$ (red symbols). The error bars are the standard error on the mean.}
\label{scatter_bin}
\end{center}
\end{figure}

In Fig. \ref{scatter_bin} a diagram with the mean scatter for all the GC sample, and for GCs brighter than K=20 and K=19 is shown. The scatter is now also represented by the median absolute deviation (\textit{mad}, indicated by squares) in addition to the $\sigma$ (represented by stars).

Both mean \textit{mad} and $\sigma$ decrease when brighter magnitude intervals are considered. This is true both for the single population and when separating in the red and blue sub-populations.
For the single population the values of $\sigma$ remain greater than the observational errors of $\sim0.4$ (see Sect. 4 in \citealt{paper1}), even for a sample with K$<$19 mag.
For the blue and red sub-populations, for the brighter magnitude samples, this is not the case any longer. The $\sigma$ is consistent with $\sim$0.4 for the first and drops below $\sim0.4$ for the latter.
However, a sample binned to brighter magnitudes will of course have a smaller mean observational error value. 
For K$<$19 mag this value is $\sim0.2$. The $\sigma$ values for K$<$19 mag are much larger than this value for the single and for the red and blue sub-populations.
Even though the \textit{mad} values are closer to $\sim0.2$ for the blue and red sub-populations, the observational errors are really $\sigma$ values and not  \textit{mad} values. Therefore there has to be something else responsible for this extra scatter.

In this section we investigate which effects other than age and metallicity could be responsible for this extra scatter, and whether these are expected to contribute significantly to any scatter.

\subsection{$\alpha-$enhanced \textit{vs.} solar-scaled tracks}
What is the effect of $\alpha-$enhancement on the integrated colours of SSPs?  For NGC\,4472 and NGC\,4486 GCs (\citealt{cohen03} and \citealt{cohen98}) [$\alpha$/Fe] estimates were derived to be $\sim +0.3$.
\cite{puzia05} finds an [$\alpha$/Fe] $\sim +0.4$ for their sample of extragalactic GCs.
In the recent study of \cite{sharina10} it is found that the [$\alpha$/Fe] ratios of GCs tend to be higher on average for giants if compared to dwarf galaxies.
If a gE has recently accreted nearby dwarfs with its GCs, they should have therefore lower [$\alpha$/Fe] ratios than the GCs formed within the protogalaxy. 
\cite{salaris07} discuss $\alpha$-enhancement issues in the study of extragalactic GCs through broad-band photometry (see their Fig. 12) and conclude that the net effect of using an $\alpha$-enhanced calibration in GCs with a solar metal mixture is to assign too young ages, especially at the reddest end. These age differences are of the order of $\sim\,2$\,Gyrs.
If on the other hand, one uses a solar scaled calibration to derive ages of $\alpha$-enhanced GCs, the ages assigned will be too old.

\begin{figure}
\begin{center}
\includegraphics[width=3.4in]{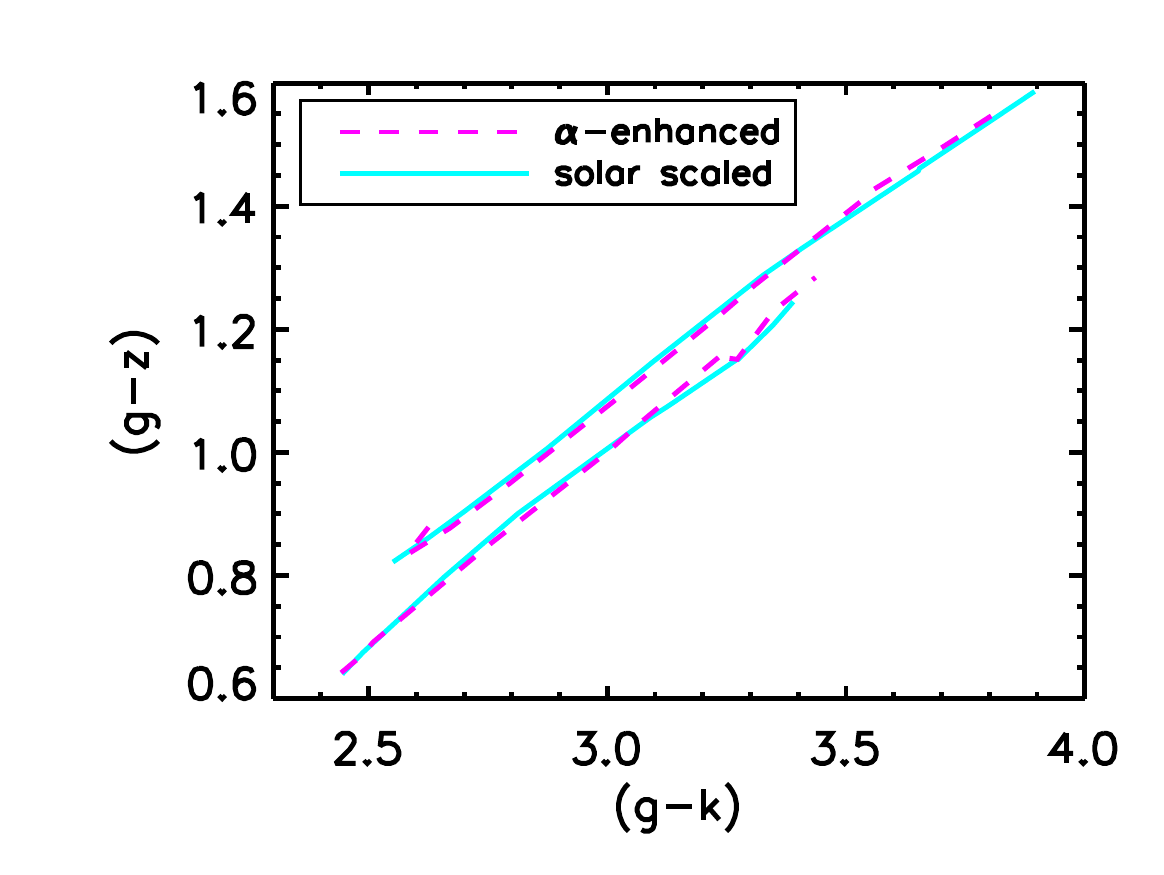} 
\caption{A 2-colour diagram with $\alpha$-enhanced and solar scaled SSPs for the ages of 2 and 10\,Gyrs. The SSPs were produced in GALEV using spectra from the BaSTI group.}
\label{2colbasti}
\end{center}
\end{figure}

Recently the BaSTI group made available $\alpha-$enhanced ([$\alpha$/Fe] $\sim +0.35$) isochrones and spectra. These were downloaded from their website and used as input in GALEV to produce $\alpha$-enhanced calibrated SSPs for the g, z and K colours.
In Fig. \ref{2colbasti} we plot 2 and 10\,Gyr SSPs in a 2-colour diagram with solar-scaled and $\alpha$-enhanced metal mixtures. The different tracks are so close together that it is very hard to tell them apart. 
The magnitude differences seen are too small to account for any noteworthy difference between the two sets of models.
The fact that \cite{salaris07} found a difference in comparing $\alpha$-enhanced and solar-scaled SSPs using V, I and K magnitudes could be attributed to the filters themselves. \cite{coelho07} found that for the majority of colour combinations the effect of $\alpha$ is in the same direction of metallicity, but in some (e.g. (B-V)) the effect is opposite to metallicity. For $(g-z)$ and $(g-K)$ the effect was unknown. 
We also tested the difference between GALEV $\alpha-$enhanced and solar scaled SSPs, produced using spectra from the BaSTI group with the V, I and K filters. The differences were slightly larger using the VIK filter combinations instead of $gzK$, confirming the results of \cite{salaris07}.
We conclude therefore that through GALEV $\alpha-$enhanced SSPs using BaSTI spectra, [$\alpha$/Fe] variations would not contribute significantly to the scatter (i.e., $\sim\,0.1-0.2$ mag) in the 2-colour diagrams using the $gzK$ filter combination, as done in this study.

\subsection{Horizontal-branch Morphology}
\begin{figure}
\begin{center}
\includegraphics[width=7cm]{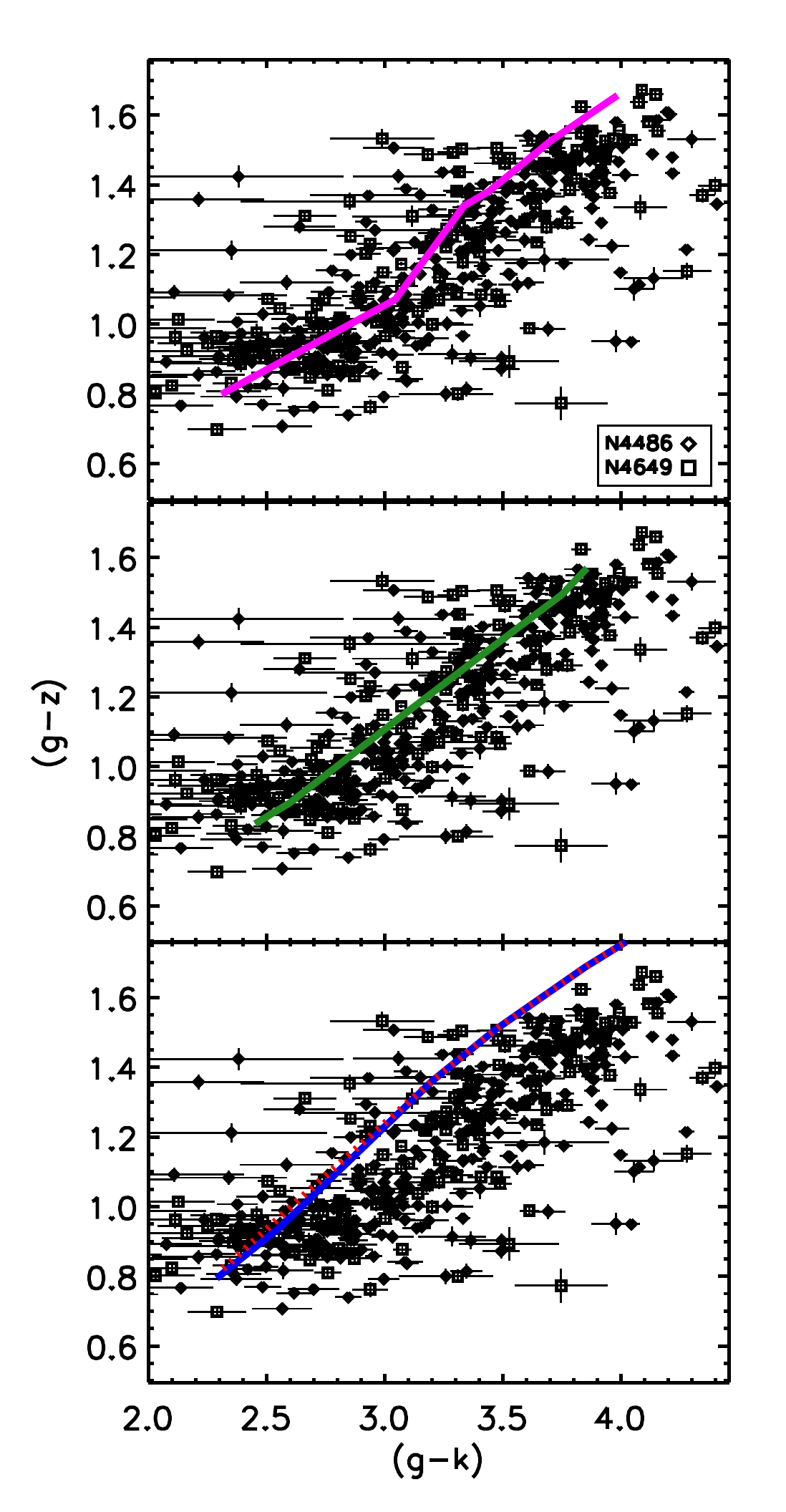}
 \caption{$(g-K)$ \textit{vs.} $(g-z)$ for GCs of NGC\,4486 and NGC\,4649, indicated by different symbols. Note the wavy feature the data presents around $(g-K)\,\sim\,3.2$ and $(g-z)\,\sim\,1.2$.
 The different panels have different SSP model over plotted: \textit{upper panel}: SpOT-Teramo 14 Gyr SSP with a realistic treatment of horizontal branch morphology (\citealt{raimondo05}),
\textit{middle panel}: Padova08 and \textit{bottom panel}: \cite{m05}.}
 \label{wiggle}
\end{center}
\end{figure}

In Fig. \ref{wiggle} the $(g-K)-(g-z)$ diagram for the joint NGC\,4486 and NGC\,4649 GC sample is shown with different models overplotted in different panels. Note the wavy feature the data presents around $(g-K)\,\sim\,3.2$ and $(g-z)\,\sim\,1.2$. A SpOT Teramo 14 Gyr SSP with a detailed treatment of HB morphology (\citealt{raimondo05}) is overplotted in the top panel. Even though this track does not fit well the data in the redder part of the diagram, it also presents the wavy feature seen in the data.
Other SSP models that do not have a detailed prescription of the HB morphology do not show this behavior (middle panel for Padova08 and bottom panel for \citealt{m05}). 
As shown in Fig. \ref{morph_scatter}, the scatter in the $\delta$ values is significantly reduced when going from a single linear relation between $(g-z)$ and $(g-K)$ to two linear relations. This separation into two linear relations simulates the effect of varying HB morphology in the 2-colour diagrams.

\subsection{Stochastic effects}
What would fluctuations in the number of bright stars do to the integrated colours of extragalactic GCs?
We have generated synthetic integrated magnitudes and colours from models
based on Padova isochrones (\citealt{marigo08}, \citealt{girardi08})
and a Monte Carlo method following \cite{sf97} in order
to assess the influence of stochastic effects on
the integrated colours of the clusters. 
Similarly as the SSPs used in Sect. 3, Padova isochrones for g, z and K were retrieved from 
CMD 2.2.
A set of isochrones combining 18 ages (2-14\,Gyrs - $\log{t}$=9.3 to 10.15 at 0.05 steps)
and 4 metallicities (Z=0.00019, 0.0006, 0.0019, 0.006) was employed.
Integrated magnitudes and colors involving g, z and K bands 
were computed by firstly distributing a prefixed
number of stars on an isochrone of given age and metallicity according
to a Salpeter initial mass function (IMF). The use of a Kroupa mass function would not make a significant difference.
The stochastic nature of the
IMF was simulated by randomly sampling stellar masses between 0.08 and
120\,M$_{\odot}$ weighted on the IMF until the preset number of stars was
reached. One hundred SSPs were simulated this way
with their integrated fluxes spread reflecting the fluctuation in the
number of bright stars.
A GC with g=23 has $\sim6\times10^5\,M\odot$ at the distance of the Virgo cluster.
This corresponds to $\sim10^6$\,stars, considering a Salpeter IMF.
For simplicity, we have considered models with a fixed number of stars.
We examine models of $10^6$ stars, since only the brightest GCs ($g\,<\,23$) are in the sample. 

\begin{figure}
\begin{center}
\includegraphics[width=8cm]{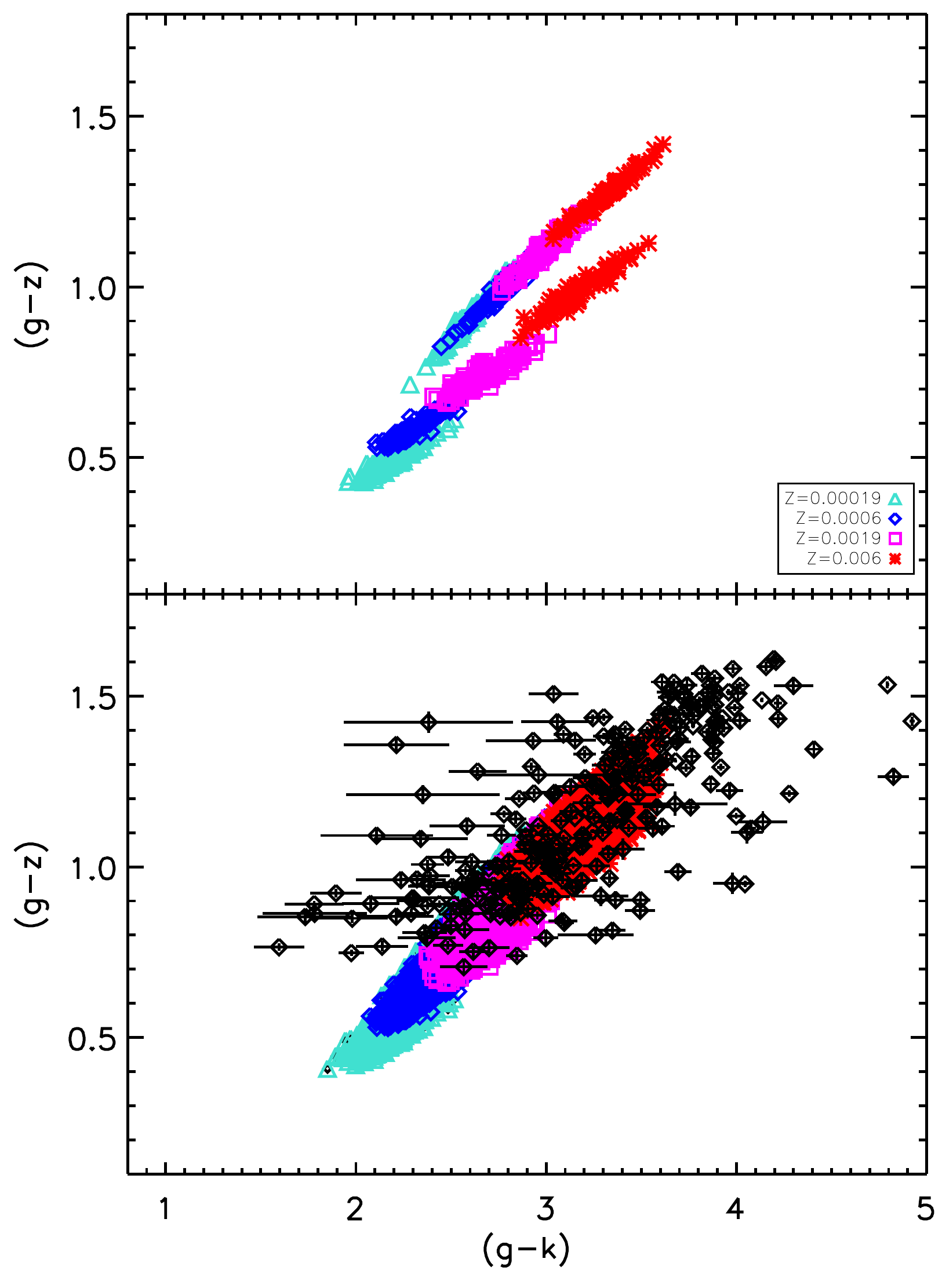} 
\caption{The synthetic $(g-K)$ \textit{vs.} $(g-z)$ diagrams for $10^{6}$ stars, for four different metallicities indicated by the different colours/symbols. \textit{Upper panel:} for the ages of 2 and 14\,Gyrs. \textit{Bottom panel}: for the age range of 2 to 14\,Gyrs and with the GCs of NGC\,4486 over plotted.}
\label{stochastic}
\end{center}
\end{figure}

In Fig. \ref{stochastic} the $(g-K) \textit{vs.} (g-z)$ synthetic diagrams are shown for $10^{6}$ stars.
The upper panel shows that the scatter on the $(g-K)$ direction due to stochastic effects produced by a 2 Gyr SSP is slightly larger than that corresponding to the older, 14 Gyr one.
The stochastic fluctuations are slightly higher for the younger SSPs if compared to the older ones due to the fact that the main sequence turnoff point occurs at a higher mass for the 2\,Gyr SSP. This gives fewer RGB/AGB stars for a given total mass due to the combination of shorter stellar lifetimes and the IMF.
The bottom panel shows that even for the combined age range of 2-14 Gyr, there is not a good match between the SSPs and the NGC\,4486 GCs.
The stochastic effects seem to be negligible as a source of colour spreading for such populous clusters.
Most of the scatter due to stochastics is orthogonal to the direction of varying age, not contributing significantly to the scatter as plotted in Figs. \ref{morph_scatter} and \ref{scatter_bin}.

\section{Discussion and implications for GC system assembly}
\begin{figure}
\begin{center}
\includegraphics[width=8cm]{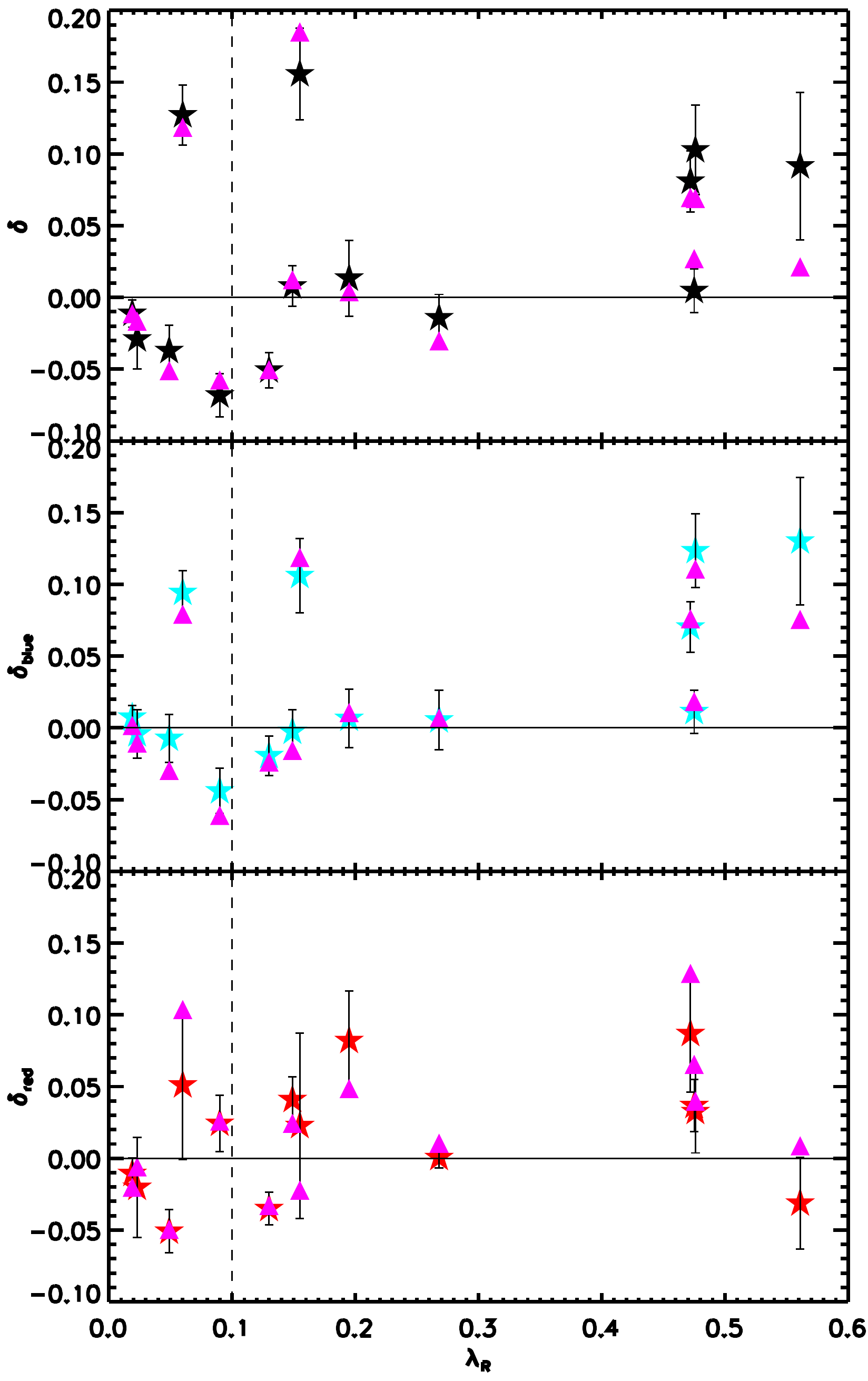} 
\caption{$\delta$, $\delta_{\rm blue}$ and $\delta_{\rm red}$ vs. $\lambda_{R}$ parameter. The dashed vertical line corresponds to the division value between slow ($\lambda_{R}<0.1$) and fast rotators ($\lambda_{R}>0.1$). The solid line and the different symbols are as indicated in Fig. \ref{morph_bluered_all}.}
\label{lambda}
\end{center}
\end{figure}

This study shows that the ages of the GC systems are correlated with the morphology of the host galaxy. Galaxies of later type are found to have a GC system with an age shift towards younger relative values if compared to galaxies of earlier type.
This finding can be crosschecked with host galaxy properties.
Major star formation events should leave imprints both in the integrated light and in the GC system of a galaxy.
Evidence of recent star formation in later-type galaxies is also found in integrated light studies.
\cite{shapiro10} combines SAURON integrated field spectroscopy of early-type galaxies with Spitzer/IRAC imaging to investigate the presence of trace star formation in such systems.
Connecting a detailed kinematic analysis given by the first data set with direct probes of star formation from Spitzer, these authors find that star formation in early-type galaxies happens exclusively in fast rotating systems.  
In Fig. \ref{lambda} $\delta$, $\delta_{\rm blue}$ and $\delta_{\rm red}$ are shown as function of $\lambda_{R}$ for the the different galaxies from \cite{emsellem07} and \cite{cap07}. This parameter is a proxy for the degree of rotation of the systems (\citealt{emsellem07}). Slow rotators have $\lambda_{R}$ below 0.1 whereas fast rotators above 0.1.
Note that the GC systems consistent with younger ages, in our picture, fall mostly among the fast rotators, with $\lambda_{R}>0.1$, with a tendency for larger $\lambda_{R}$ values ($>$\,0.45) for $\delta$ and $\delta_{\rm blue}$.
Slow rotators on the other hand seem to have a population of GCs consistent with $\delta$ and $\delta_{\rm blue}$ $<0$, implying old ages. 
This is suggestive of a more recent GC formation-connected to recent star formation in the host galaxy only occurring in fast rotating systems.
The two outlying cases, with low $\lambda_{R}$ values but high $\delta$ and $\delta_{\rm blue}$ values are NGC\,4406 ($\lambda_{R}<0.1$) and NGC\,4382 ($\lambda_{R}>0.1$).  
There is evidence that the fast rotator NGC\,4382 could have an artificially low measured $\lambda_{R}$ in virtue of its significant inclination (see \citealt{emsellem07}).

Star formation occurring later in time in a galaxy is expected to come from metal-enriched gas.
Therefore, one would presume that the metal-rich clusters would be driving this relation, yet this does not appear to be the case.
The blue population appears to drive the relation between GC age and host galaxy morphological type. In this context, the blue GCs of the E5/S0s/SAB(0) are younger. 
This is surprising and contrary to the view that the metal-poor GCs formed (exclusively) at high redshift (\citealt{bs06}).
We have investigated whether this trend could be caused by
systematics in the data such as seeing, aperture corrections,
sky subtraction, etc., and conclude that this is not the case.
There is, however a trend of anti-correlation with the number of clusters in a galaxy. However, this only reflects the fact that later type galaxies (S0's) have less clusters than earlier type ones (E0's). Furthermore, a trend between $\delta$ values and the brightness of the cluster is not present.  
Here, the direction of $\delta$ values shifts are interpreted as age shifts.
However, given our limited understanding of (the necessary ingredients to construct) SSP models there is still the possibility that this is not an age effect.
It could be perhaps due to an unknown outcome of a certain group of stars in the clusters (e.g. HB, AGB).

A possibility is that the blue population drives the relation between GC age and host galaxy morphological type as a result of minor mergers. There would exist two possible ways for assembling these younger GCs.
Dwarf galaxies have a more extended star formation history than galaxies of larger mass (\citealt{tolstoy09}) and contain almost exclusively metal-poor GCs (\citealt{forbes00}).  
One possibility is that these blue GCs had formed in the accreted dwarf galaxies (\citealt{cote98}).
Another possibility is that the GCs were formed during the merger event from the gas reservoir of accreted dwarfs, presumably with metal-poor gas.
Evidence for this latter possibility is found in \cite{muratov10}. In this model for GC formation in the hierarchical framework of galaxy assembly, mergers of smaller hosts create exclusively blue clusters and mergers of more massive galaxies create both red and blue clusters. 
It remains uncertain, however, whether such effects would be expected
to operate more prominently in S0-type galaxies compared to 
ellipticals. It is also unclear to what extent the age distributions
of blue GCs in S0-type galaxies are directly related to the presence 
of recent star formation in the discs as revealed by SAURON. Unlike
the blue GCs, the discs tend to be metal-rich, and spatially our
blue GCs do not appear to be associated with the discs either.

In the Milky Way, there is substantial evidence that part of its GC system is associated with accreted dwarf galaxies.
The age-metallicity relation study of \cite{fb10} for galactic GCs suggests that there are two tracks of objects in the Galaxy. One with constant old ages $\sim13$\,Gyrs, spanning a large metallicity range $-2.2<$[Fe/H]$<-0.6$, possibly formed in situ. The second track, branches to younger ages and the objects appear clumped at a mean age value of $\sim$\,10.5 Gyrs and intermediate metallicity, [Fe/H]\,$\sim$-1.3. This latter track seems to be dominated by GCs associated with dwarf galaxies accreted by the Galaxy.
A $\delta$ $\sim$0.05-0.1 corresponds in very rough terms to an age shift of $\sim$2-8\,Gyrs. Such an age shift is therefore comparable to the difference in age between GCs formed in situ and GCs accreted from dwarf galaxies in the Milky Way. 
Moreover the [Fe/H] value ($\sim$-1.3) of this clump of objects, associated to dwarf galaxies, is closer to the metal-poor peak metallicity value than to the metal-rich one of GC systems of early-type galaxies. For example, the metal-poor peak of NGC\,4406 is at [Fe/H]$=-1.23$, whereas the metal-rich one at [Fe/H]$=-0.56$ (\citealt{bs06}, based on the empirical relation of \citealt{peng06}). 
Therefore if this younger clump of objects of the Galaxy was to be placed in NGC\,4406, for example, it would belong to its metal-poor population.
This is suggestive of younger GCs possibly belonging to the metal-poor population of a galaxy.
In this context the age and metallicity of the blue population of clusters are related to the objects associated to dwarf galaxies.

There is also a trend of later type galaxies exhibiting a greater positive $\delta_{\rm red}$ shift than earlier type ones (bottom panel of Fig. \ref{morph_bluered_all}), although with much more scatter than for $\delta_{\rm blue}$. The maximum shift occurs for E5 galaxies.
A great $\delta_{\rm red}$ shift could be indicative of the host galaxy having experienced a major merger. 
From the \cite{az92} scenario, formation of metal-rich GCs would occur during the merger of disc galaxies. 
Alternatively, it is tempting to link the large $\delta_{\rm red}$ values to the amount of "disc" formation that happened in a particular galaxy (\citealt{kunt06}).
For example, the fast rotators with large $\delta_{\rm red}$ values, NGC\,4660, NGC\,4473 and NGC\,4526 have a prominent Mg$b$ enhanced, metal-enriched central disc-like structure (\citealt{kunt06}, \citealt{kunt10}).    

For all this to get beyond speculation, further investigation is required. 
Increasing the number of measured $\delta$'s for different GC system would perhaps strengthen (or weaken) the relation between GC system age and morphological galaxy type.
Spectroscopy of GCs in later type early-type galaxies (S0's, perhaps E5's) might give it further confirmation. 
Moreover, cosmological simulations up to GC system formation and assembly could give further constraints on the nature of the mechanisms that shape the age distributions of GCs in present day early-type galaxies.

This study also shows that the age distribution of the GC system of NGC\,4365 is not different from other similar type galaxies (e.g. NGC\,4486 and NGC\,4649).
The optical colour distribution of this GC system is indeed different from others. This does not imply however, significant differences in the ages, but rather in the metallicity distributions.

\section{Summary and Conclusions}
Using multi-band g, z and the newly derived K-band photometry we have compared differentially the GC systems in a sample of 14 E/S0 galaxies. We have also studied the effect of different parameters in various SSP models. The results are outlined below.
\begin{enumerate} 
\item The most recent Padova SSPs for old ages show less of an offset with respect to the photometry than previously published models. 
However, a formal fit still yields intermediate ages.
\item The age distribution of the GC system of NGC\,4365 appears similar to that of other large ellipticals in the sample.
Although its colour distribution is different from other systems in $(g-z)$, this difference is orthogonal to the direction of decreasing/increasing age shift in the $(g-K)$-$(g-z)$ plane.
\item When the Padova08 SSPs integrated colours are used as SEDs the fits of the median ages of the blue and red sub-populations with AnalySED are found to be consistent with a single age, within one sigma. This makes them consistent with a great range of age values (from 1 to 14 Gyrs), with the exception of some S0 galaxies.
\item A differential comparison without using SSP models is performed. This comparison is done in the direction of age shift, traced by the quantity $\delta$. A correlation between galaxy morphological type and the mean delta is found. Galaxies of earlier type host genuinely old GCs and galaxies of later type, younger on average clusters. The blue population appears to be driving this relation. The relation for the red population is not as clear as for the blue one.
\item The $(g-K)$ vs. $(z-K)$ parameter space is analysed and it is concluded that there is a non-linear feature around $(g-K)\sim3.2$ and $(z-K)\sim2$. This feature is suggested to be due to the HB morphology. It is also present in SpOT-Teramo SSPs which have a detailed modelling of this stellar evolution phase.
\item The influence of stochastic effects is investigated to address the spread seen in the 2-colour planes. It is concluded that the observational uncertainties are larger than the influence of stochastic effects for the cluster mass relevant to our study. In fact, the observational uncertainties can account for most of the observed scatter when considering blue and red clusters as separate populations.
\item We find that there is residual scatter in the delta parameter
  at bright magnitudes that cannot be accounted for by 
  observational errors. This may indicate an intrinsic scatter
  in the GC ages of up to several Gyr.
\end{enumerate}

\begin{acknowledgements}
We thank the referee for comments that improved the paper.
We acknowledge Paula Coelho for interesting discussions.
JFCSJ acknowledges the Brazilian institution FAPEMIG (grant APQ-00117/08).
This research has made use of the NASA/IPAC Extragalactic Database (NED) which is operated by the Jet Propulsion Laboratory, California Institute of Technology, under contract with the National Aeronautics and Space Administration
\end{acknowledgements}


\begin{thebibliography}{}

\bibitem[Acosta Pulido et al.(2003)]{apulido03}Acosta Pulido, J. A., Ballesteros, E., Barreto, M., et al., 2003, ING Newsl., 7, 15 
\bibitem[Anders et al. (2004)]{anders04}Anders, P., Bissantz, N.. Fritze-v. Alvensleben, U. \& de Grijs, R., 2004, MNRAS, 347, 196
\bibitem[Ashman \& Zepf(1992)]{az92}Ashman, K.M., Zepf, S,E., 1992, ApJ, 384, 50
\bibitem[Ashman et al.(1994)]{abz94}Ashman, K. M., Bird, C. M. \& Zepf, S. E., 1994, AJ, 108, 2348
\bibitem[Brodie \& Strader(2006)]{bs06} Brodie J.P. \& Strader J. 2006, ARA\&A, 44, 193
\bibitem[Cenarro et al.(2007)]{cenarro07} Cenarro,A.J., Beasley,M.A.,Strader,J., Brodie, J.P \& Forbes, D.A. 2007, AJ,134,391
\bibitem[Cappellari et al.(2007)]{cap07} Cappellari, M., Emsellem, E., Bacon, R. et. al. 2007, MNRAS, 379, 418
\bibitem[Cohen et al.(1998)]{cohen98} Cohen, J. G., Blakeslee, J. P. \& Ryzhov, A., 1998, ApJ, 496, 808
\bibitem[Cohen et al.(2003)]{cohen03} Cohen, J. G., Blakeslee, J. P. \& C\^ot\'e, 2003, ApJ, 592, 866
\bibitem[C\^ot\'e et al.(1998)]{cote98} C\^ot\'e, P., Marzke, R.O., West, M.J., 1998, ApJ, 501, 554
\bibitem[Coelho et al.(2007)]{coelho07}Coelho, P., Bruzual, G., Charlot et al., 2007, MNRAS, 382, 498
\bibitem[Charlot \& Bruzual(in prep)]{cb07}Charlot, S \& Bruzual, G, in prep
\bibitem[Chies-Santos et al.(2010a)]{paper1} Chies-Santos, A. L., Larsen, S. S., Wehner, E. M., Kuntschner, H., Strader, J., Brodie, J. P., \textit{Paper I, A\&A accepted, arXiv1010.0687}  
\bibitem[Elmegreen \& Efremov(1997)]{ee97}Elmegreen, B. G. \& Efremov, Y. N., 1997, ApJ, 480, 235
\bibitem[Emsellem et al.(2007)]{emsellem07}Emsellem, E., Cappellari, M., Krajnovi\'c, D., et al., 2007, MNRAS, 379, 401
\bibitem[Ferrarese et al.(2006)]{ferrarese06}Ferrarese, L.,  C\^ot\'e, P.,Jord\'an, A. et al., 2006, ApJS, 164, 334
\bibitem[Forbes et al.(2000)]{forbes00}Forbes, D. A., Masters, K. L., Minniti, D., Barmby, P., 2000, A\&A, 358, 471
\bibitem[Forbes et al.(1997a)]{forbes97}Forbes, D. A., Brodie, J. P, \& Huchra, J., 1997 AJ, 113, 887
\bibitem[Forbes et al.(1997b)]{fbg97}Forbes, D.A., Brodie, J. P. \& Grillmair, C. J., 1997, AJ, 113, 1652
\bibitem[Forbes \& Bridges(2010)]{fb10}Forbes, D. A., Bridges, T., 2010, MNRAS, 404, 1203
\bibitem[Girardi et al.(2008)]{girardi08}Girardi, L., Dalcanton, J., Williams, B. et al. \textit{the ANGST/ANGRRR Collaboration}, 2008, PASP 120, 583
\bibitem[Goudfrooij et al.(2001a)]{goud01spec}Goudfrooij, P., Mack, J., Kissler-Patig, M., Meylan, G. \& Minniti, D., 2001, MNRAS, 322, 643
\bibitem[Goudfrooij et al.(2001b)]{goud01}Goudfrooij, P., Alonso, M. V., Maraston, C. \& Minniti, D., 2001, MNRAS, 328, 237
\bibitem[Hempel et al.(2003)]{hempel03} Hempel, M., Hilker, M., Kissler-Patig, M., Puzia, T. H., Minniti, D. \& Goudfrooij, P., 2003 A\&A, 405, 487
\bibitem[Hempel et al.(2007a)]{hempel07}Hempel, M., Zepf, S., Kundu, A., Geisler, D., \& Maccarone, T. J., 2007, ApJ, 661, 768
\bibitem[Hempel et al.(2007b)]{hempel07AA}Hempel, M., Kissler-Patig, M., Puzia, T. H., \& Hilker, M., 2007, A\&A, 463, 493
\bibitem[Kotulla et al.(2008)]{kotulla08}Kotulla, R., Fritze, U. \& Anders, P., 2008, MNRAS, 387, 1149
\bibitem[Kuntschner(2000)]{kunt00}Kuntschner, H., 2000, MNRAS, 315, 184
\bibitem[Kuntschner et al.(2006)]{kunt06}Kuntschner, H., Emsellem, Eric., Bacon, R. et al., 2006, MNRAS, 369, 497
\bibitem[Kuntschner et al.(2010)]{kunt10}Kuntschner, H., Emsellem, E., Bacon, R., et al., 2010, \textit{MNRAS, accepted}
\bibitem[Larsen(1999)]{larsen99}Larsen,S.S.,1999, A\&AS, 139, 393
\bibitem[Larsen \& Brodie(2000)]{lb00}Larsen, S. S., Brodie, J. P., 2000, AJ, 120,2938
\bibitem[Larsen et al.(2001)]{larsen01}Larsen,S.S, Brodie, J.P., Huchra, J.P., Forbes, D.A.,Grillmair, C.J., 2001, AJ, 121, 2974
\bibitem[Larsen et al.(2005)]{lbs05}Larsen, S. S., Brodie, J. P., Strader, J., 2005, A\&A, 443, 413
\bibitem[Lee et al.(1994)]{lee94}Y. W. Lee, P. Demarque, R. Zinn, 1994, ApJ, 423, 248
\bibitem[Manchado et al.(2004)]{manchado04}Manchado, A., Barreto, M., Acosta-Pulido et al., 2004, SPIE, 5492, 1094
\bibitem[Maraston(2005)]{m05}Maraston, C., 2005, MNRAS, 362, 799
\bibitem[Marigo et al.(2008)]{marigo08}Marigo, P., Girardi, L., Bressan, A. et al., 2008, A\&A, 482, 883
\bibitem[Muratov \& Gnedin(2010)]{muratov10}Muratov, A. L. \& Gnedin, O. Y., 2010, ApJ, 718, 1266
\bibitem[Peng et al.(2006)]{peng06}Peng, E. W., Jord\'an, A. C\^ot\'e, P. et al., 2006, ApJ, 639, 95
\bibitem[Piotto et al.(2007)]{piotto07}Piotto, G., Bedin, L. R.; Anderson, J.; et al. 2007, ApJ, 661, 53
\bibitem[Puzia et al.(2002)]{puzia02}Puzia, T. H., Zepf, S. E., Kissler-Patig, M., Hilker, et al., 2002, A\&A, 391, 453
\bibitem[Puzia et al.(2004)]{puzia04}Puzia, T. H., Kissler-Patig, M., Thomas, D. et al. 2004, A\&A, 415, 123
\bibitem[Puzia et al.(2005)]{puzia05}Puzia, T.H., Kissler-Patig, M., Thomas, D., et al. 2005, A\&A, 439, 997
\bibitem[White(1978)]{white78}White, S. D. M., 1978, MNRAS, 184, 185
\bibitem[Worthey(1994)]{worthey94}Worthey, G., 1994, ApJS, 95, 107
\bibitem[Raimondo et al.(2005)]{raimondo05}Raimondo, G., Brocato, E., Cantiello, M. \& Capaccioli, M., 2005, AJ, 130, 2625 
\bibitem[Salaris \& Cassisi(2007)]{salaris07}Salaris, M. \& Cassisi, S., 2007, A\&A, 461, 493
\bibitem[Santos \& Frogel(1997)]{sf97} Santos Jr., J. F. C., Frogel, J. A. 1997, ApJ 479, 764
\bibitem[Shapiro et al.(2010)]{shapiro10}Shapiro, K. L., Falc\'on-Barroso, J. van de Ven, G. et al., 2010, MNRAS, 402, 2140 
\bibitem[Sharina et al.(2010)]{sharina10}Sharina, M. E., Chandar, R., Puzia, T. H., Goudfrooij, P., Davoust, E.,  2010, MNRAS, 405, 839
\bibitem[Springel et al.(2005)]{springel05}Springel, V., White, S. D. M., Jenkins, A. et al.  2005, Nature, 435, 629
\bibitem[Stetson(1987)]{stetson87}Stetson, P. B., 1987, PASP, 99, 191
\bibitem[Strader et al.(2005)]{strader05}Strader,J., Brodie, J.P., Cenarro,A.J.,Beasley,M.A. \& Forbes, D.A. 2005, AJ,130,1315
\bibitem[Strader et al.(2007)]{strader07}Strader J., Beasley, M. A. \& Brodie, J. P., 2007, AJ, 133, 2015 
\bibitem[Tolstoy et al.(2009)]{tolstoy09}Tolstoy, E., Hill, V., Tosi, M., 2009, ARA\&A, 47, 371
\bibitem[Thomas et al.(2005)]{thomas05}Thomas, D., Maraston, C., Bender, R., Mendes de Oliveira, C., 2005, ApJ, 621, 0673
\bibitem[Yamada et al.(2006)] {yamada06}Yamada, Y., Arimoto, N., Vazdekis, A. \& Peletier, R. F., 2006, ApJ, 637, 200
\bibitem[de Zeeuw et al.(2002)]{zeeuw02}de Zeeuw, P. T., Bureau, M., Emsellem, E. et al., 2002, MNRAS, 329, 513


\end{thebibliography}
\end{document}